\documentclass[a4paper,fleqn,usenatbib]{mnras}
\usepackage{newtxtext,newtxmath}
\usepackage[T1]{fontenc}
\usepackage{ae,aecompl}
\usepackage{amssymb,multirow,color, amsmath}
\usepackage{dcolumn}% Align table columns on decimal point
\usepackage{bm}% bold math
\usepackage{amssymb,amsmath,amsfonts,graphicx,epsf}
\usepackage{color}
\usepackage{enumerate}
\newcommand{\bvec}[1]{\mathbf{#1}}
\newcommand{\pa}{\partial}
\newcommand{\dfdxi}{\frac{\partial F}{\partial \xi}}
\newcommand{\dtwofdxi}{\frac{\partial^2 F}{\partial \xi^2}}
\newcommand{\dfdmu}{\frac{\partial F}{\partial \mu}}

\title{Kinetic Theory and Fast Wind Observations of the Electron Strahl}

\author[K. Horaites et al.]{
Konstantinos Horaites,$^{1}$\thanks{E-mail: horaites@wisc.edu}
Stanislav Boldyrev,$^{1,2}$
Lynn B. Wilson III,$^{3}$
Adolfo F. Vi\~nas,$^{3}$
\newauthor and Jan Merka$^{3,4}$
\\
$^{1}$Department of Physics, University of Wisconsin -- Madison, 1150 University Avenue, Madison, WI 53706, USA\\
$^{2}$Space Science Institute, Boulder, Colorado 80301, USA\\
$^{3}$NASA Goddard Space Flight Center, Greenbelt, Maryland, USA\\
$^{4}$University of Maryland, Baltimore County, Goddard Planetary Heliophysics Institute, Baltimore, United States
}

\date{Accepted XXX. Received YYY; in original form ZZZ}

\pubyear{2017}

\begin{document}
\label{firstpage}
\pagerange{\pageref{firstpage}--\pageref{lastpage}}
\maketitle

\begin{abstract}
We develop a model for the strahl population in the solar wind -- a narrow, low-density and high-energy electron beam centered on the magnetic field direction.   Our model is based on the solution of the  electron drift-kinetic equation at heliospheric distances where the plasma density, temperature, and the magnetic field strength decline as power-laws of the distance along a magnetic flux tube. Our solution for the strahl depends on a number of parameters that, in the absence of the analytic solution for the full electron velocity distribution function (eVDF), cannot be derived from the theory. We however demonstrate that these parameters can be efficiently found from matching our solution with observations of the eVDF made by the Wind satellite's SWE strahl detector. The model is successful at predicting the angular width (FWHM) of the strahl for the Wind data at 1 AU, in particular by predicting how this width scales with particle energy and background density. We find the strahl distribution is largely determined by the local temperature Knudsen number $\gamma \sim |T dT/dx|/n$, which parametrizes solar wind collisionality. We compute averaged strahl distributions for typical Knudsen numbers observed in the solar wind, and fit our model to these data. The model can be matched quite closely to the eVDFs at 1 AU; however, it then overestimates the strahl amplitude at larger heliocentric distances. This indicates that our model may be improved through the inclusion of additional physics, possibly through the introduction of ``anomalous diffusion'' of the strahl electrons.
\end{abstract}

%{\em Introduction.}---
\section{Introduction} The strahl is a narrow, magnetic field-aligned population of suprathermal electrons \citep{rosenbauer77} routinely observed in the ambient solar wind. The strahl is comprised of ``thermal runaway'' electrons \citep{gurevichistomin79}. Because high energy particles are relatively insensitive to Coulomb collisions ($\nu_{coll}\sim v^{-3}$), electrons of sufficiently high energy can stream over large distances without coming into local thermal equilibrium \citep{scudderolbert79}. The strahl electrons generally move antisunward, because the relatively hot inner regions of the heliosphere act as a source of high-energy particles. However, in some instances---notably, during the transit of interplanetary coronal mass ejections---``counterstreaming'' strahls \citep[e.g., ][]{gosling87,anderson12}, which are directed towards the sun, are also observed. When it is prominent, the strahl can provide the dominant contribution to the electron heat flux \citep{pilipp87}, which is an important source of heating of the solar wind during its non-adiabatic expansion \citep[e.g., ][]{stverak15}.

The beam-like shape of the strahl in velocity space is believed to come from the competition of two physical processes: the mirror force and pitch angle scattering. The mirror force narrows this population, so that electron velocities run nearly parallel to the local magnetic field. The runaway electrons see a weakening field as they travel to larger heliocentric distances, which converts the particles' perpendicular velocity into parallel velocity. The observed strahls, however, are not as narrow as one would expect from conservation of the first adiabatic invariant. It is therefore inferred that a scattering process provides some diffusion that broadens the distribution. 

This pitch angle scattering is usually attributed to either Coulomb collisions or wave-particle interactions, and interest in the latter has been partially motivated by the perceived failure of the former. In particular, measurements conducted by \cite{lemonsfeldman83} showed the strahl to be even broader than would be predicted by incorporating Coulomb collisions into exospheric theory, and it was suggested that a source of  ``anomalous diffusion'' in the form of wave-particle interactions may be scattering the strahl particles.  

Nonetheless, there is a wealth of evidence indicating that on average, characteristics of the strahl can be strongly correlated with the Coulomb collisionality of the background solar wind. We here refer to ``collisionality'' in terms of the Knudsen number $\gamma(x=const.) \propto T^2 / n$, which parameterizes the ratio between advective and Coulomb scattering terms in the kinetic equation.
Prominent, narrow strahls have long been associated with the fast solar wind \citep[e.g., ][]{feldman78, pilipp87, anderson12}. The fast wind is less collisional on average, since typically $n$ is lower and $T$ higher than in the slow wind. \cite{ogilvie00} specifically noted that prominent, narrow strahls can be seen when the fast solar wind has very low density. \cite{salem03} showed that the solar wind heat flux $q$, which comes from the skewness of $f(\bvec{v})$ owing partially to the strahl, is correlated with $\gamma$. \cite{bale13} clarified this picture, demonstrating in a broad statistical study of Wind data that the solar wind heat flux in fact scales with $\gamma$ exactly as predicted by classical theory \citep{spitzerharm53} in the collisional regime $\gamma \ll 1$. In the regime $\gamma \gtrsim 1$, the heat flux was observed to ``saturate'' at a collisionless value $q\sim n v_{th} T$. This saturation is also predicted, and may owe to the onset of various plasma instabilities \citep[e.g, ][]{cowiemckee77}. In \cite{horaites15}, average field-parallel cuts of the electron distribution computed from Helios data were shown to vary with $\gamma$ as predicted by a ``self-similar'' kinetic theory, with strahl amplitude increasing with $\gamma$. The fact that the solar wind data are so well-ordered by $\gamma$ is a strong indication that Coulomb collisions play a central role in the physics that shapes the electron strahl.

Most wave-particle theories of strahl scattering have considered the effect of whistler waves \citep[e.g., ][]{vocks03, vocks05, saitogary07}. These waves can resonate with the cyclotron motion of strahl electrons, scattering the particles and broadening their distribution. \cite{seough15} suggested a variation on this mechanism, in which anti-sunward halo particles {\it not} scattered by whistlers can be focused by the magnetic field into the strahl. Whistlers may be generated by the electron heat flux instability \citep[e.g., ][]{gary75, gary94}; this view has been supported by studies of whistler wave events \citep[e.g, ][]{wilson13, stansby16}. Whistlers may also be excited by the temperature anisotropy of the strahl population \citep[e.g., ][]{vinas10}, or may help comprise the small-scale interplanetary turbulence \citep[e.g., ][]{pagel07, boldyrev13}. Recently, the first direct observations of long-lived (lasting longer than 5 minutes) whistler waves were made \citep{lacombe14}, using Cluster data.  A subsequent study \citep{kajdic16} demonstrated that the presence of whistler waves is correlated with broader strahl widths. However, the authors emphasized that whistler waves were detected infrequently by Cluster in the ``pristine'' (unperturbed by transient structures) fast wind: only 37 time intervals sustaining whistlers for a minute or longer were observed in a 10-year period. Looking beyond whistler waves, \cite{pavan13} proposed that strahl distributions focused by the mirror force should generate Langmuir oscillations. In their numerical simulations, quasilinear oscillations continually scatter the strahl electrons, developing a steady state in which scattering balances magnetic focusing. %Observational evidence was provided by \cite{gurgiolo12} that wave-particle interactions may scatter strahl particles; however, the authors did not .

The detailed shape of the strahl distribution, which carries information about the physics that formed it, has been characterized in multiple observational studies. Using data from the Imp 6, 7, and 8 satellites, \cite{feldman78} found that the angular breadth of the strahl distribution decreases monotonically with energy in the fast wind. This is consistent with the predictions of Coulomb scattering models \citep[see e.g., ][]{saitogary07, fairfieldscudder85} and some wave scattering models \citep[e.g., ][]{pavan13}.
%This corroborates the view that the strahl is mediated by Coulomb collisions, which are less effective for high energy electrons \citep[see e.g., ][]{saitogary07}.
%Using Helios 1 and 2 data, \citep{pilipp87} provided a comprehensive summary of the basic features of the electron. 
This picture is not always observed in the data, however; \cite{anderson12} reported that the strahl width can either increase or decrease with energy. Some instances where the strahl broadens with energy have been associated with rare transient events, such as solar electron bursts \citep{dekoning07} and periods of increased wave activity \citep{pagel07}.
Counter to the notion that the magnetic field continuously focuses the strahl as the particles travel away from the sun, \cite{hammond96} found that the strahl width (at a given energy) actually increases with heliocentric distance $r$ for Ulysses data $r>$1 AU (see also \cite{graham17}).
% Prominent, narrow strahls have been associated with the fast wind, and with regions of low particle density \citep{fitzenreiter98, ogilvie00}. 

The goal of this work is to develop an analytical model for the electron strahl population in the solar wind. The accurate analytic derivation of the full electron distribution function is a very complicated task since it requires one to solve the electron kinetic equation in an expanding background, subject to the boundary conditions at the base of the solar wind and at infinity. In our approach we build upon our previous work \cite[][]{horaites15}, and consider the electron kinetic equation in the range of distances where the parameters of the solar wind (density, temperature, magnetic field strength) can be approximated as power laws of the distance along a magnetic flux tube. We then expect that the solution for the strahl is {\em scale invariant}, namely, it has similar structure at different distances after appropriate rescaling of the velocity field and the amplitude of the distribution function. This allows one to find a nontrivial solution for the electron strahl, which incorporates the effects of advection, focusing by the expanding magnetic field, and pitch angle scattering by Coulomb collisions.\footnote{As it would greatly complicate our current project, we do not yet incorporate the effects of wave-particle scattering in our model.}

The scale-invariance does not allow us to determine the solution uniquely, which reflects the fact that we do not solve a full boundary value problem, and, therefore, do not have enough information to fully describe the fast electrons that stream from the hot solar surface. We however demonstrate that the remaining arbitrariness can be removed very efficiently by matching our analytic solution with the observations. %, assuming scale invariance of the distribution.       

As a test of our theory, we examine high resolution measurements of the strahl distribution made by the Wind satellite, whose Solar Wind Experiment (SWE) had an electrostatic analyzer dedicated to measuring the strahl distribution \citep{ogilvie95}. Wind's strahl detector had very fine ($\sim$4.5$^\circ$) angular resolution, making it ideal for measuring the shape of the strahl distribution. Following previous observational papers \citep[e.g.][]{ogilvie00}, we will characterize the breadth of the distribution in terms of the angular full width at half maximum (FWHM). We will demonstrate in section \ref{obs_sec} that the strahl widths are accurately described by our theory, as it correctly predicts how the width decreases with particle energy and increases with background density.

Our conclusions differ somewhat from those of \cite{lemonsfeldman83}, in that we do not find the strahl widths at 1 AU to be too broad, necessarily. In our model, the parameters defining the large-scale structure of the solar wind (variation of temperature, density, etc.) are intertwined with the predicted shape of the strahl distribution, including its width, energy-dependence, and variation with heliocentric distance. So although in section \ref{fmodel_fave_sec} we will show that our model can be matched to average eVDFs at 1 AU, in section \ref{discussion_sec} we will infer from published measurements of the solar wind structure that the strahl should increase in amplitude with distance, relative to the core. This may be in contradiction with measurements made by \cite{maksimovic05} and \cite{stverak09}, which show the opposite trend, at least in the low-energy portion of the strahl distribution. We believe a more complete theory of the strahl may require the inclusion of additional physics, for instance in the form of anomalous diffusion. The search for such a theory should be well-constrained by the existing body of observations, including those presented here. 
%{\bf We find that.... and that the theory using only Coulomb scattering may not describe adequately all the properties of the strahl. Our analysis points to the necessity to include the scattering provided by interactions with turbulence...?} 

\section{Theory}\label{theory_sec}
Consider a solar wind in which the electron temperature $T$, density $n$, and magnetic field strength $B$ all vary as power laws with distance $x$ along a flux tube: $n=n_0 (x/x_0)^{\alpha_n}$, $T=T_0 (x/x_0)^{\alpha_T}$, and $B = B_0 (x/x_0)^{\alpha_B}$. Throughout this paper, the subscript $0$ will be used to note values of a variable at reference position $x=x_0$. Specifically, we will let $x_0\sim 1 AU$ be the position of the Wind spacecraft. We assume that the number of particles in the nearly maxwellian core of the electron distribution function is typically much larger than the number of particles in the strahl (and in the halo),  so the electron temperature is mostly defined by the core population \citep[e.g.][]{stverak09}.  Without loss of generality, we will express the electron distribution in the form:
\begin{eqnarray}
\label{f_F_eq}
f(\bvec{v}, x) ={N F(\bvec{v}/v_{th}(x), x})/{{T(x)}^\alpha},
\end{eqnarray}
where $v_{th} \equiv \sqrt{2T/m_e}$ is the electron thermal speed and $N$ is a constant set by the normalizations $\int f({\bf v}, x) d^3v= n(x)$, $\int F({\bf u}, x) d^3u =1$. Applying these normalizations to equation~(\ref{f_F_eq}) leads to the relation  
\begin{eqnarray}
\alpha = 3/2 - \alpha_n / \alpha_T .
\end{eqnarray} 

We also define the Knudsen number $\gamma(x)$, which parametrizes the Coulomb collisionality:
\begin{eqnarray}
\label{gamma_eq}
\gamma(x) \equiv -T^2 ({d\ln T}/{dx})/(2 \pi e^4 \Lambda n) = \lambda_{mfp}/L_T,
\end{eqnarray}
and the electric Knudsen number $\gamma_E(x)$ which is the ratio of the electric field to the Dreicer field
\begin{eqnarray}
\label{gamma_e_eq}
\gamma_E(x) \equiv {E_\parallel e T}/({2 \pi e^4 \Lambda n}),
\end{eqnarray}
where $E_\|$ is the component of the electric field parallel to the magnetic field.
In these expressions we use the Coulomb logarithm $\Lambda$, which we treat as a constant ($\Lambda = 25.5$), and introduce the electron mean free path
\begin{eqnarray}
\lambda_{mfp} \equiv T^2 /(2 \pi e^4 \Lambda n),
\end{eqnarray}
and the length scale of temperature variation
\begin{eqnarray}
L_T \equiv ({d\ln T}/{dx})^{-1} = -x/\alpha_T.
\end{eqnarray}
%The strahl electron population forms  

At frequencies much smaller than the electron gyrofrequency, and at scales much larger than the electron gyroscale, the distribution function $f(v,\mu, x)$ obeys the drift-kinetic equation \citep[e.g., ][]{kulsrud83}, which, in a steady state ($\partial f/\partial t=0$), takes the form:
\begin{eqnarray}
\label{drift_kinetic_eq_udzero}
\mu v \frac{\pa f}{\pa x} &-& \frac{1}{2}\frac{d\ln B}{dx}v (1-\mu^2) \frac{\pa f}{\pa \mu} - \nonumber \\
& -&\frac{e E_\parallel}{m} \left[\frac{1-\mu^2}{v} \frac{\pa f}{\pa \mu}  + \mu \frac{\pa f}{\pa v}\right] = \hat C(f),
\end{eqnarray}
where $v=|{\bf v}|$,  $\mu = \bvec{v}\cdot\hat B/v$, the unit vector $\hat B$ points anti-sunward along the local magnetic field. Equation (\ref{drift_kinetic_eq_udzero}) presumes the $\bvec{E} \times \bvec{B}$ drift is negligible. It describes the evolution of the (gyrotropic) distribution function in the reference frame of a fixed flux tube, for electrons with speeds much greater than the solar wind speed: $v\gg v_{sw}$. 

The effect of Coulomb collisions are described by the operator $\hat C(f)$.  We will be mostly interested in the strahl component of the distribution function, which has a relatively low density and a relatively high velocity as compared to the core distribution. Under these conditions, one can use the linearized, high-energy form of the collision integral \cite[e.g.,][]{helandersigmar02}:
\begin{eqnarray}\label{coll_op_eq}
\hat C(f)  = \frac{4 \pi n e^4 \Lambda}{m_e^2} \left[\frac{\beta}{v^3}\frac{\pa}{\pa \mu} (1-\mu^2)\frac{\pa f}{\pa \mu}  \right. \nonumber \\
   +\left. \frac{1}{v^2}\frac{\pa f}{\pa v} + \frac{v_{th}^2}{2v^2}\left( -\frac{1}{v^2}\frac{\pa f}{\pa v} + \frac{1}{v}\frac{\pa^2 f}{\pa v^2}\right) \right] .
\end{eqnarray}
In this expression, $\beta \equiv (1 + Z_{eff})/2$, and $Z_{eff}$ is the effective charge of the ion species \citep[e.g., ][]{rathgeber10}. We note $Z_{eff}\approx 1$ in the solar wind, where most ions are protons. The derivation of Eq.~(\ref{coll_op_eq}) requires an explanation. The collision integral (\ref{coll_op_eq}) describes the interaction of fast ``test" electrons with the Maxwellian electron and ion ``field" populations (core), which are assumed to have similar temperatures $T_e \sim T_i$. In the derivation of Eq.~(\ref{coll_op_eq}) it is assumed that the density of the test particles is much smaller than the density of the core particles and the energy of the test particles is much larger than the core energy, $(v/v_{th})^2 \gg 1$. The first term in the rhs of Eq.~(\ref{coll_op_eq}) describes the pitch angle scattering of the test electrons by the core electrons and ions, while the remaining terms describe the energy exchange of the test electrons with the core electrons. Due to low density of the test electrons, their mutual interactions are neglected. We will naturally identify the test electrons with the strahl electrons, while the field particles will correspond to the nearly maxwellian core of the solar wind distribution function. 

Substituting expression (\ref{f_F_eq}) into equation (\ref{drift_kinetic_eq_udzero}) and introducing the dimensionless variable $\xi \equiv (v/{v_{th}})^2$, we find the following equation for $F(x, \xi, \mu)$:
\begin{eqnarray}
\label{ssk_eq}
&{}&\lambda_{mfp}(x)\frac{\partial F}{\partial x}\nonumber \\
 &+&\gamma(x) \left[\alpha \mu F + \mu\xi\dfdxi + \frac{\alpha_B}{2}(\alpha + 1/2)(1-\mu^2)\dfdmu \right] \nonumber \\
&-& \gamma_E(x) \left[ \mu \dfdxi + \frac{1 - \mu^2}{2\xi}\dfdmu \right]=\hat C(F),
\end{eqnarray}
where the collision operator (\ref{coll_op_eq}) takes the form \citep[][]{krasheninnikov88,krasheninnikovbakunin93}: 
\begin{eqnarray}
\label{self_similar_collision_operator_eq}
\hat C(F)=\frac{1}{\xi} \left[\dfdxi + \dtwofdxi \right] + \frac{\beta}{2 \xi^2} \frac{\pa}{\pa \mu} \left(1 - \mu^2\right) \dfdmu .
\end{eqnarray}
Equation (\ref{ssk_eq}) is nearly identical to the self-similar kinetic equation studied in \cite{horaites15}, where it was assumed that the distribution function $F$, and the Knudsen numbers $\gamma(x)$ and $\gamma_E(x)$ are independent of~$x$. In the present work we do not make such simplifying assumptions, and allow for explicit dependence of the distribution function $F$ on $x$. We however assume that similarly to density, temperature, and magnetic field, $\gamma(x)$ varies along a magnetic flux tube as a power law: 
\begin{eqnarray}
\gamma(x) =\gamma_0 (x/x_0)^{\alpha_\gamma},
\end{eqnarray} 
where, as it follows from the definition~(\ref{gamma_eq}), 
\begin{eqnarray}
\alpha_\gamma = 2 \alpha_T - \alpha_n - 1.
\end{eqnarray}

We now apply the kinetic equation (\ref{drift_kinetic_eq_udzero}) to the strahl electron population. The field-aligned, high-energy strahl population corresponds to $\mu \approx 1$, $\xi \gg 1$. Let us look for asymptotic solutions to equations (\ref{ssk_eq}) and (\ref{self_similar_collision_operator_eq}) in this regime. We approximate $(1-\mu^2) \approx 2(1-\mu)$, and neglect the ambipolar electric field terms and the collisional energy-exchange terms (the terms containing $\xi$-derivatives in (\ref{self_similar_collision_operator_eq})), which are higher-order in $1/\xi$. 

Neglecting the electric field terms in equation (\ref{ssk_eq}) amounts to assuming $\gamma_E \ll \gamma \xi$, which requires further justification based on an estimate of the function $\gamma_E(x)$. For a rough estimate of the function $\gamma_E(x)$ we note that in spherically symmetric exospheric models of the solar wind, in which the magnetic field lines are assumed to point radially ($x=r$), the ambipolar electric field $E_\parallel(x)$ can be evaluated from the electron momentum equation \citep[e.g., ][]{hollweg70, jockers70}:  %see also Lemaire & Scherer 1971
\begin{equation}
e n E_{\parallel} + \frac{d}{dx}(n T) = 0.
\end{equation}
This equation retains only the most important terms (neglecting, e.g., gravity) and assumes that the electron bulk flow is subsonic. In a scale-invariant model in which $n(x)$ and $T(x)$ are power laws, the solution to this equation can be approximated as:
\begin{equation}
E_{\parallel}(x) \sim \frac{T(x)}{e x}.
\end{equation}
For such a model, we obtain using Eqs.~(\ref{gamma_eq}) and (\ref{gamma_e_eq})   that $\gamma_E(x) / \gamma(x) $$\sim$1.  It is therefore reasonable to assume $\gamma_E \ll \gamma \xi$, and we can indeed neglect the electric field terms in (\ref{ssk_eq}). 
 We also note that the latter condition may be re-written as $m_ev^2\gg eE_\|L_T$. Noting that $L_T$ is on the order of the electron mean-free-path $\lambda_{mfp}$, we see that the electric field does not affect the dynamics of the fast electrons between the collisions.  
%We also note that the onset of the strahl population is known to occur at a ``breakpoint'' energy $\xi > 1$ \citep{feldman75, scudder79}.
%We note also that it has previously been predicted that the strahl can only form at energies above a so-called ``breakpoint'' $\xi \gg 1$ \citep{scudderolbert79, pilipp87}.
%Thus we see that if the condition $\gamma_E \ll \gamma \xi$ is satisfied for given $\xi$ at a particular $x$, it should 

Defining 
\begin{eqnarray}
\kappa \equiv 2 - \alpha_\gamma / \alpha_T,
\end{eqnarray}
and
\begin{eqnarray} 
\alpha^\prime \equiv \kappa- (\alpha+1/2)\alpha_B, 
\end{eqnarray}
we write the resulting equation for $F(x, \xi, \mu)$ in the form:
\begin{eqnarray}\label{strahl_asympt_ssk_eq}
\lambda_{mfp}(x) \frac{\pa F}{\pa x} &+& \gamma(x) \left\{\alpha F + \xi \frac{\pa F}{\pa \xi} + (\kappa-\alpha^\prime) (1-\mu) \frac{\pa F}{\pa \mu}\right\} = \nonumber \\
&=&\frac{\beta}{\xi^2}\frac{\pa}{\pa \mu} (1-\mu) \dfrac{\pa F}{\pa \mu}.
\end{eqnarray}
Equation (\ref{strahl_asympt_ssk_eq}) includes the effects of advection, angular focusing due to the spatially expanding magnetic field, and pitch angle scattering due to $e-e$ and $e-p$ Coulomb collisions.  To bring equation (\ref{strahl_asympt_ssk_eq}) into a more tractable form, we would like to use the variable $v = \sqrt{\xi}v_{th}(x)$ instead of the variable $x$. This choice of variables leads to the following equation for $F(v,\xi,\mu)$:
\begin{equation}\label{strahl_asympt_ssk_eq_newvars}
\alpha F + \xi \frac{\pa F}{\pa \xi} + (\kappa - \alpha^\prime)(1-\mu)\frac{\pa F}{\pa \mu} = \frac{\beta}{\gamma(x) \xi^2} \frac{\pa}{\pa \mu} (1-\mu) \frac{\pa F}{\pa \mu}.
\end{equation}
In this equation the Knudsen number $\gamma(x)$ should be expressed as a function of $\xi$ and $v$ as follows
\begin{eqnarray}
\gamma(x)\equiv \gamma_0\left(v/v_{th,0} \right)^{2\alpha_\gamma/\alpha_T}\xi^{-\alpha_\gamma/\alpha_T}.
\end{eqnarray}
We show this intermediate step to demonstrate that equation (\ref{strahl_asympt_ssk_eq_newvars}) contains no derivatives with respect to $v$, which significantly simplifies the PDE. As a result, our solutions will include arbitrary functions of $v$, which, as we demonstrate later, may be effectively determined from comparison with observations. 

To find the solution of Eq.~(\ref{strahl_asympt_ssk_eq}) we perform another change of variables by treating $F$ as a function of independent variables $v$, $\eta$, and $z$, where $\eta =\ln \xi$ and $z= \xi^\kappa (1-\mu)$. The differential equation for $F(v, \eta, z)$ then takes the form:
\begin{equation}\label{strahl_asympt_ssk_eq_newvars2}
\alpha F + \frac{\pa F}{\pa \eta} + \Big[\alpha^\prime z - \frac{\beta}{G(v)} \Big]\frac{\pa F}{\pa z} = \frac{\beta}{G(v)} z \frac{\pa^2 F}{\pa z^2},
\end{equation}
where 
\begin{eqnarray}\label{g_v_eq}
G(v) = \gamma_0\left(v/v_{th,0} \right)^{2\alpha_\gamma/\alpha_T} =  \gamma(x)\xi^{(\alpha_\gamma/\alpha_T)}.
\end{eqnarray}
%$G(v) = G_0 v^{(2\alpha_\gamma/\alpha_T)} =  \xi^{(\alpha_\gamma/\alpha_T)} \gamma$ and $G_0$ is a constant. 
The solution of this advection-diffusion equation can be sought in the form $F(v, \eta, z) = \exp\{-z a(v, \eta) + b(v, \eta)\}$, where $a(v, \eta)$ and $b(v, \eta)$ are general functions. Substituting this solution form into equation (\ref{strahl_asympt_ssk_eq_newvars}), and equating coefficients at like powers of $z$, we obtain the following two equations:
\begin{eqnarray}
\label{z_sep_eq1}
\alpha + \frac{\pa b}{\pa \eta} + \Big(\frac{\beta}{G(v)}\Big) a = 0, \\
\label{z_sep_eq2}
\frac{\pa a}{\pa \eta} + \alpha^\prime a + \Big(\frac{\beta}{G(v)}\Big) a^2 = 0.
\end{eqnarray}

Equation (\ref{z_sep_eq2}) can be solved for $a(v, \eta)$ by integration, and $b(v,\eta)$ is similarly obtained from equation (\ref{z_sep_eq1}). Substituting these functions into our assumed solution form yields the strahl distribution $F(v, \xi, \mu)$:
\begin{equation}\label{asymptotic_strahl_exact_eq}
F(v, \xi, \mu) = C_1(v) \frac{\exp\left\{\frac{\alpha^\prime C_2(v) \xi^{\kappa-\alpha^\prime}(1-\mu)}{1 + \beta C_2(v) \xi^{-\alpha^\prime}/G(v)}  \right\}} {  {1 + \beta C_2(v) \xi^{-\alpha^\prime}/G(v)}} \xi^{-\alpha},
\end{equation}
where $C_1(v)$ and $C_2(v)$ are arbitrary functions. Due to assumed scale invariance of the strahl distribution function, we expect that the physically relevant solution should correspond to either $1\ll \beta C_2(v)\xi^{-\alpha^\prime}/G(v)$ or $1\gg \beta C_2(v)\xi^{-\alpha^\prime}/G(v)$. 
It may be argued the observed breadth of the strahl is more consistent with the first case (see Appendix), in which case equation (\ref{asymptotic_strahl_exact_eq}) reduces to the relatively simple expression:
%It may be argued that observational angular broadening of the strahl with the temperature is more consistent with the first case (see Appendix), in which case equation (\ref{asymptotic_strahl_exact_eq}) reduces to the relatively simple expression:
\begin{equation}\label{asymptotic_strahl_eq}
F(v, \xi, \mu) = C(v)\xi^{\alpha^\prime - \alpha}\exp\left\{\tilde{\gamma}(v, \xi) \Omega \xi^2 (1-\mu) \right\},
\end{equation}
where $C(v)$ is an arbitrary (dimensionless) function of~$v$. For convenience later in the paper, we here introduced the notation 
\begin{eqnarray}
\tilde{\gamma} \equiv  T^2/(2\pi e^4 \Lambda n x) = -\gamma(x) / \alpha_T,
\end{eqnarray}
and 
\begin{eqnarray}\label{omega_def}
\Omega \equiv -\alpha^\prime \alpha_T/\beta.
\end{eqnarray}
For relevant solar wind profiles, $\alpha^\prime$ and $\alpha_T$ (and $\Omega$) are negative. 

Equation (\ref{asymptotic_strahl_eq}) is an exact solution to equation (\ref{strahl_asympt_ssk_eq_newvars}). The field-parallel ($\mu=1$) energy profile of the strahl amplitude depends on the function $C(v)$, which in turn determines how the strahl amplitude varies with distance $x$ at {\em fixed} $\xi$. The presence of the arbitrary function $C(v)$ reflects the fact that the fast electrons are not generated locally, but rather produced at the base of the solar wind and then propagate along the magnetic field lines almost without collisions. This function could be obtained if the analytic solution for the full electron distribution function (for all energies and distances) were available. Such an exact solution is however very complicated and currently cannot be obtained. The function $C(v)$ can however be well constrained from the analysis of the observational data. We will do this in the next section, where we compare equation (\ref{asymptotic_strahl_eq}) with direct measurements of the distribution function made by Wind's strahl detector.

\section{Observations}\label{obs_sec}
\subsection{SWE Strahl Detector}
The Wind satellite's strahl detector is a toroidal electrostatic analyzer \citep{ogilvie95}, which directly sampled the solar wind electron distribution function (eVDF) at a heliocentric distance $ r=$1 AU. The instrument's 12 anodes are set in a vertical pattern in a plane that contains the spacecraft spin axis, spanning a field of view $\pm28^\circ$ centered around the ecliptic. Wind's spin axis is set at a right angle with the ecliptic plane, allowing different azimuthal angles to be sampled as the spacecraft spins (3 sec spin period). %what is the spin period?
These azimuthal bins have a fixed separation of 3.53$^\circ$. Each strahl distribution measured by the spacecraft consists of a 14x12 angular grid of electron counts, that was measured at a fixed energy during a single spacecraft spin. 
Counts can be converted into physical units of $f(\bvec{v})$ (e.g., cm$^{-6}$s$^{3}$) in the standard fashion by accounting for the detector efficiency and geometric factor. Accompanying each strahl measurement is an analogous 14x12 measurement of the ``antistrahl'', made at a clock angle 180$^\circ$ with respect to the strahl measurement. The detector voltage was set to a different value each spin, so that 32 energies from 19.34 to 1238 eV would be sampled in as many rotations.

In the original mode of operation, each measurement grid was centered on the nominal average Parker spiral (in the ecliptic plane, $45^\circ$ offset from the radial direction $\hat r$). In practice, however, the local magnetic field only fell within the field of view of the detector about half the time. This prompted a revision of the instrument software in February 1999 \citep{ogilvie00}, which matched the clock angle of the strahl measurement with the instantaneous measurement of the magnetic field provided by Wind's Magnetic Field Investigation (MFI).

Our data set ranges from January 1, 1995 to May 30, 2001, which nearly covers the operational lifetime of the strahl instrument. The strahl detector was reconfigured shortly after this period to serve as a replacement for SWE's Vector Electron/Ion Spectrometer (VEIS), whose power supply had recently failed. % (A.~F. Vi\~nas 2016, personal communication).   %\citep{vinas_comm16}. 

\subsection{Strahl Data}\label{strahl_data_sec}%cleaning method
The data studied here are derived primarily from Wind's SWE strahl detector, supplemented by plasma data from SWE/VEIS and vector magnetic field data from MFI \citep{ogilvie95, lepping95}. The background electron density $n$ is taken from the VEIS measurement of proton density $n_p$, which is a reasonable estimate for the quasineutral, proton-dominated solar wind. This is more reliable than direct measurement of $n$ with the electron instrument since measurements of the proton distribution are less susceptible to spacecraft charging effects \citep[see, e.g., ][]{deforest72, pilipp87}. The temperature is calculated from the second-order moment of the eVDF measured by VEIS. We note that this measurement overestimates the core temperature $T$ by a small factor (10-20$\%$), since the suprathermal populations are included in the calculation. %(citep someone, or use TNR or Stuart's core fits, or conduct SWE fits). 
Plasma parameters (which are used to calculate $\xi$, $\gamma$, etc.) from other instruments are associated with each strahl measurement by matching to the nearest measurement time. %say this more clearly?
If plasma data are not available within 5 minutes of the strahl measurement, then the data are excluded from our analysis. We also exclude data for which the $\hat B$ direction, as measured by the MFI instrument, is outside of the strahl detector's field of view.

Each measurement made by the SWE strahl detector is a composition of signals from the strahl, halo, and (at low energies) core components of the electron distribution \citep[e.g., ][]{pilipp87}. In order to conduct a statistical study of the fast wind strahl, we developed an automated procedure for isolating the strahl signal from the background formed by the other populations. The procedure is as follows:

%note that we don't correct for spacecraft potential, high energies, negligible effect

%\begin{enumerate}
\begin{description}
\item {\it Find the strahl}---For each 14x12 strahl distribution $f_s$ (measured at a single energy), find the bin where the distribution is at a maximum, and designate the nominal velocity direction of electrons measured in that bin as the ``peak direction''.\\
\item {\it Remove the halo}---Let the maxima of the measured strahl and associated anti-strahl distributions be designated $f_{s,max}$, $f_{a,max}$ respectively. Zero out (ignore in future analysis) the bins of the strahl distribution where the criterion $f_s < Max \{f_{a,max}\times3/2, f_{s,max}/5\}$ is satisfied. \\
\item {\it Clean up residual noise}---Calculate the pitch angle $\theta$, relative to the peak direction, of every bin in the 14x12 grid. 
%The pitch angle between two bins is the angle between the bins' nominal viewing directions.
Find the minimum pitch angle $\theta_{min} = Min\{\theta\}$ among the bins zeroed out in the previous step. Zero out every bin in the strahl distribution $f_s$ that satisfies $\theta > \theta_{min}$. From this point forward, ``$f_s$'' will refer to the cleaned strahl distribution resulting from the above procedure.
%\end{enumerate}
\end{description}

\begin{figure}
\includegraphics[width=\linewidth]{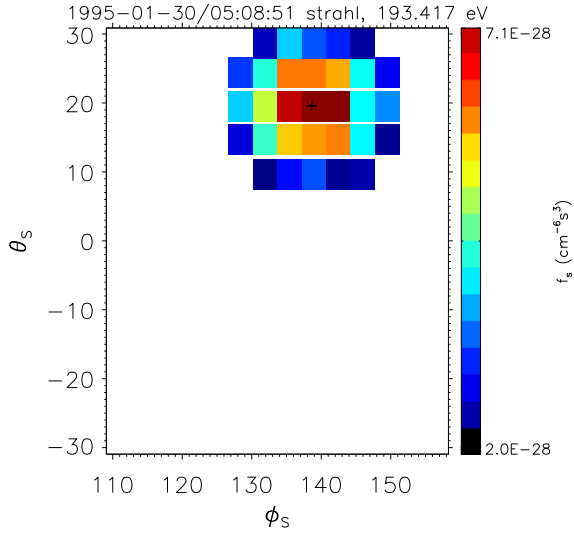}
\caption{\label{strahl_spectrum_fig} An example strahl spectrum $f_s$ (linear z-axis scale), after applying an automated procedure for removing the halo population. This plot can be compared with figure 3, \protect\cite{fitzenreiter98}, which shows the ``raw'' spectrum. The variables $\theta_S$, $\phi_S$ are spherical (GSE) coordinates that describe the velocity direction of the measured electrons. The angle $\theta_S=0$ corresponds with the ecliptic. Note that the detector's 12 anodes are not evenly spaced in $\theta_S$. The magnetic field direction $\hat B$ is determined through the process of nonlinear fitting described in section \ref{width_obs_sec}, and is shown in this example by a ``$+$'' symbol.}% with a significant gap between anodes near the $\theta_S=0$, so that the vertical axis of this plot is somewhat distorted.}
\end{figure}

\noindent This cleaning procedure leads to a very clearly defined strahl, an example of which is shown in figure \ref{strahl_spectrum_fig}.

For strahl distributions measured after the February 1999 software revision, anomalously high count rates were observed when the sun was in the detector's view (R.~J. Fitzenreiter 2016, personal communication).    %\citep{fitzenreiter_comm16}. 
These spurious counts were caused by photoelectrons, and should be removed from our analysis. As a simple correction, we zero out (prior to step ``Find the strahl'' above) data from all 12 anodes at a given azimuthal angle, if one of these anodes pointed within 10$^\circ$ of the sun's position.

\subsection{Analysis of Strahl Widths}\label{width_obs_sec}

We now compare the SWE strahl detector data with our model. We will only consider the fast solar wind, i.e. when the solar wind bulk velocity exceeded 550 $km/sec$. The strahl detector measures each distribution at fixed energy $\xi$ and position $x$. We therefore predict that the distribution function $F(\mu)$ should fall off exponentially with $(1-\mu)$, as it follows from equation \ref{asymptotic_strahl_eq}:

\begin{equation}\label{f_mu_eq}
F(\mu) \propto \exp \{\tilde{\gamma} \Omega \xi^2 (1-\mu)\}.
\end{equation}

\noindent Let us express equation \ref{f_mu_eq} in terms of the pitch angle with respect to the magnetic field, $\theta \equiv \cos^{-1} (\mu)$. That is, for the small angles ($\theta \approx 0$) relevant to the strahl, we can approximate $\mu \approx 1-{\theta^2}/2$, implying $F$ should fall off as a Gaussian with $\theta$:

\begin{equation}\label{f_theta_eq}
F(\theta) \sim \exp \Big\{ \frac{\tilde{\gamma} \Omega \xi^2 \theta^2}{2} \Big\}.
\end{equation}

\noindent This prediction agrees with previous attempts to model the strahl; functions of the form $f(\theta)\propto \exp(-C\theta^2)$, where $C$ is some constant, have been used to fit measured strahl distributions at fixed energy \citep{hammond96, anderson12}.

The full width at half maximum of the strahl, $\theta_{FWHM}$, follows immediately from equation \ref{f_mu_eq}:
 
\begin{equation}\label{fwhm_eq}
\begin{split}
\theta_{FWHM} &= 2\cos^{-1}\Big[{1-\frac{\ln(2)}{\tilde{\gamma} |\Omega| \xi^2}}\Big] \\
              &\approx \frac{360}{\pi \xi} \sqrt{\frac{2 \ln (2)}{\tilde{\gamma} |\Omega|}} deg.
\end{split}
\end{equation}

\noindent The approximate expression in equation \ref{fwhm_eq}, which we arrived at by substituting $\mu \approx 1-{\theta^2}/2$ into equation \ref{f_mu_eq}, reveals how $\theta_{FWHM}$ should scale with other locally observable quantities, i.e. density $n$, core temperature $T$, energy $\mathcal{E}=m_e v^2/2$, and distance $x$ along the flux tube. In terms of these physical quantities, equation \ref{fwhm_eq} can be expressed:

\begin{equation}\label{fwhm_eq_local}
%\theta_{FWHM} \approx 16.6 \sqrt{\frac{n[cm^{-3}] r[AU]}{|\Omega|\mathcal{E}[eV]^2}} radians
\theta_{FWHM} \approx 951 \sqrt{\frac{n x}{|\Omega|\mathcal{E}^2}} deg.
\end{equation}

\noindent Equation \ref{fwhm_eq_local} is written with the choice of units, $[n]=cm^{-3}$, $[x]=AU$, $[\mathcal{E}] = eV$. In the inner heliosphere, we can make the reasonable estimate $x\approx r$ (for radial field lines). 
%In the context of our model, $\Omega$ is a constant that can be inferred by measuring $\theta_{FWHM}$ at known $n$, $r$, and $\mathcal{E}$. 

In the context of our model, $\Omega$ is a constant that depends on solar wind structure (eq. \ref{omega_def}). 
From equation \ref{fwhm_eq}, we obtain the following scaling relations for fixed $x$ and $\Omega$:

\begin{enumerate}[i]
\item For given $n$, $\theta_{FWHM}\propto \mathcal{E}^{-1}$
\item For given $\mathcal{E}$,  $\theta_{FWHM}\propto \sqrt{n}$
%\item For $\mu=1$, $F \propto \xi^{\epsilon}$
\end{enumerate}

\noindent The local quantities $n$ and $\mathcal{E}$, which determine the breadth of the strahl in our model, are known to high accuracy. Measurements of these parameters have relative errors 10$\%$ and 3$\%$ for $n$ and $\mathcal{E}$, respectively \citep{ogilvie95}. We note also that our prediction for $\theta_{FWHM}$ is independent of $T$.

To test our prediction for $\theta_{FWHM}$, we fit each strahl distribution $f_s$ to our model (\ref{f_mu_eq}). 
Recalling $f_s(\mu) \propto F(\mu)$ (see eq. (\ref{f_F_eq})), let us define $z \equiv \ln (f_s/f_{s,max})$, $y \equiv (1-\mu)$, $m\equiv \tilde{\gamma} \Omega \xi^2$, and write equation \ref{f_mu_eq} as:

\begin{equation}\label{z_y_eq}
 z(y) = my + \mathcal{Z},  % for $\xi=const$.. 
\end{equation}

\noindent where $m$ and $\mathcal{Z}$ are constants.

Since our model (\ref{z_y_eq}) is linear when expressed in these variables, it would seem natural to employ the weighted ordinary least squares (OLS) technique to find the parameters $m$, $\mathcal{Z}$ for each distribution. This could be accomplished by fitting to the data $\overline{z}_i$ comprising the distribution, which is measured at independent coordinates $y_i$ (``$i$'' indexes the bins of $f_s$). However, it turns out that there is considerable error in our determination of the magnetic field direction $\hat B$ which must be corrected for. If our measurement of $\hat B$ (as determined by the MFI instrument) is off by even a few degrees, errors in $y_i$ will lead to significant inaccuracies in the determination of the strahl width. This effect is most problematic when the angular error of the $\hat B$ direction is on the order of $\theta_{FWHM}$, which can occur for narrow strahls.

We therefore conduct a weighted {\it nonlinear} least squares fit to the distribution function ($z$), in which the direction $\hat B$ is determined by the fitting procedure. Our fits are generated using the MPFIT software \citep{markwardt09}, which implements the Levenberg-Marquardt algorithm. We fit to the (2D) model function:

\begin{equation}\label{2d_model_function}
% z(\phi_S, \theta_S)=m y(\phi_S, \theta_S; \phi_B, \theta_B) + \mathcal{Z}.
 z(\phi_S, \theta_S)=m y(\phi_S, \theta_S; \phi_B, \theta_B) + \mathcal{Z}.
\end{equation}

\noindent  Here we introduced the function $y(\phi, \theta; \phi^\prime, \theta^\prime)$,

\begin{equation}
y(\phi, \theta; \phi^\prime, \theta^\prime) = 1 - \hat U(\phi, \theta) \cdot \hat U(\phi^\prime, \theta^\prime),
\end{equation}

\noindent and $\hat U(\phi, \theta)$ is a function that produces a unit vector pointing in the direction specified by $\phi$, $\theta$ (azimuth and altitude, in GSE coordinates). 

%\noindent This is the same as the OLS function (\ref{z_y_eq}) described above, with some nuance. Here there are two independent coordinates, $\phi_S$ and $\theta_S$, which represent azimuth and altitude in spherical (GSE) coordinates. The model (eq. \ref{2d_model_function}) has four fit parameters: $m$, $\phi_B$, $\theta_B$, and $\mathcal{Z}$. The parameters $m$ and $\mathcal{Z}$ are as above. The direction $\hat B$ is specified by the fit parameters $\phi_B$ and $\theta_B$, which represent the azimuth and altitude of $\hat B$ in spherical (GSE) coordinates. The $y_i$ data are interpreted as above, but now $y_i=1-\mu_i$ depends on the $\phi_{S}$, $\theta_{S}$ identified with the $i^{th}$ bin's nominal look direction, as well as on the fit parameters $\phi_B$ and $\theta_B$. Namely, if $\hat S_i$ is the unit vector representing the velocity direction of electrons measured in the $i^{th}$ bin of the detector, then $y_i = 1- \hat S_i(\phi_S, \theta_S) \cdot \hat B(\phi_B, \theta_B)$.

Equation (\ref{2d_model_function}) is the same as the OLS function (\ref{z_y_eq}) described above, with some nuance. Now there are two independent coordinates, $\phi_S$ and $\theta_S$, which represent azimuth and altitude in spherical (GSE) coordinates. The model (eq. \ref{2d_model_function}) has four fit parameters: $m$, $\phi_B$, $\theta_B$, and $\mathcal{Z}$. The parameters $m$ and $\mathcal{Z}$ are as above. The direction $\hat B$ is specified by the fit parameters $\phi_B$ and $\theta_B$, i.e. $\hat B = \hat U(\phi_B, \theta_B)$. % which represent the azimuth and altitude of $\hat B$ in spherical (GSE) coordinates.
 The $y_i$ data are interpreted as above, but now $y_i=1-\mu_i$ depends on the $\phi_i$, $\theta_i$ identified with the nominal velocity direction of electrons measured in the $i^{th}$ bin, as well as on the fit parameters $\phi_B$ and $\theta_B$. Namely, $y_i = y (\phi_i, \theta_i; \phi_B, \theta_B)$.
%*** introduce \hat S earlier, then will be clearer to understand throughout ***

\begin{figure}
\includegraphics[width=1\linewidth]{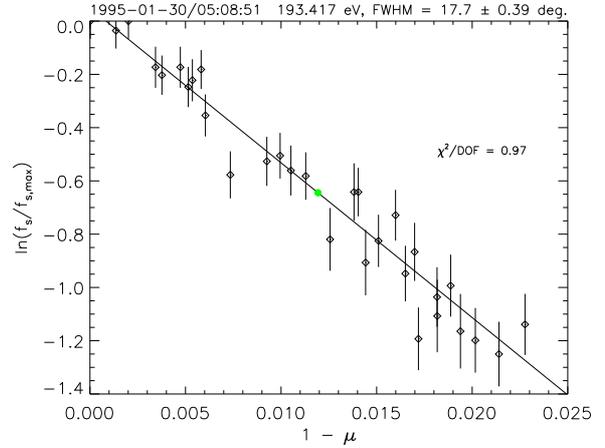}
\caption{\label{fit_strahl_width_fig} 
An illustration of our fitting method: the angular distribution $f_s$ (from Figure \ref{strahl_spectrum_fig}) is displayed above as a pitch angle distribution $z(y)$, where $y = (1-\mu)$ and $z = \ln(f_s/f_{s,max})$. The data in the vicinity of the strahl peak is fit to the function $z = m y + \mathcal{Z}$. Our method of fitting, a nonlinear least squares fit that allows the $\hat B$ direction to vary, is a weighted fit that accounts for the uncertainties $\sigma_i$ of the $\overline{z}_i$ measurements. The $\sigma_i$ are shown above as error bars. The full width at half-maximum of the strahl (green dot), $\theta_{FWHM}$, is calculated from $m$ according to equation \ref{fwhm_m_eq}. }
\end{figure}

Our weighted fit requires an estimate of the standard error of the $\overline{z}_i$ measurements, which we denote as $\sigma_i$. This can be estimated by assuming the strahl detector obeys Poisson (``counting'') statistics. We assume the raw number of counts $\zeta_i$ registered by the detector in the $i^{th}$ bin is sufficiently large, so that we can approximate the error of $\zeta_i$ as Gaussian-distributed, with standard deviation $\sqrt{\zeta_i}$. Then, we find $\sigma_i\approx 1/\sqrt{\zeta_i}$ from straightforward error propagation (noting $f_s$ is proportional to counts).

 Our fitting procedure minimizes the chi-squared statistic: 

\begin{equation}
\chi^2 = \sum_{i=1}^N (z_i-\overline{z}_i)^2/\sigma_i^2. 
\end{equation}
%In this sum, the model function $z$ is evaluated at $z_i$. 

\noindent Here, $(z_i - \overline{z}_i)$ represents the difference between our model function and the data, for the $i^{th}$ bin of the distribution. $N$ is the number of non-zero data points in the 12$\times$14 strahl distribution $f_s$; we only conduct a fit if there are at least 6 points left after applying the cleaning procedure (section \ref{strahl_data_sec}), so $6\leq N\leq168$. 
 Figure \ref{fit_strahl_width_fig} shows an example of our fitting procedure, applied to the data appearing in figure \ref{strahl_spectrum_fig}. 

The normalized chi-squared statistic $\chi^2/DOF$ can be used to test goodness of fit. Here $DOF=N$-4 represents the degrees of freedom of our 4-parameter model function (\ref{2d_model_function}). For our fast wind data, we find $\langle \chi^2/DOF\rangle=$ 1.12. The angle brackets indicate the average value among the $>$100,000 fits in our analysis.
To calculate this average, we excluded outlier fits for which $\chi^2/DOF>10$, which constituted only $1\%$ of the completed fits. Since $\langle\chi^2/DOF\rangle\approx 1$, we conclude that equation \ref{f_mu_eq} accurately describes the strahl data. %, i.e. $\ln F \propto (1-\mu)$ at constant $\xi$.

Having fit for the slope $m$ for each strahl distribution $f_s$, we calculate the distribution's ``measured $\theta_{FWHM}$'': 

\begin{equation}\label{fwhm_m_eq}
\theta_{FWHM} = 2 \cos^{-1}\{1-\log(1/2)/m\}.
\end{equation}

\noindent This formula follows from equation (\ref{z_y_eq}), utilizing the definitions of $y$ and $z$. We can additionally calculate a ``measured $\Omega$'':

\begin{equation}\label{omega_m_eq}
\Omega = \frac{m}{\tilde{\gamma} \xi^2}.
\end{equation}

\noindent For completeness, we note a selection criterion: a distribution $f_s$ was only retained for study if the fit for the slope $m$ (eq. \ref{2d_model_function}) yielded a ``measured $\theta_{FWHM}$'' (eq. \ref{fwhm_m_eq}) that was less than twice the maximum pitch angle among the bins of $f_s$. This avoids extrapolation errors in our determination of $\theta_{FWHM}$.

\begin{figure}
\includegraphics[width=1\linewidth]{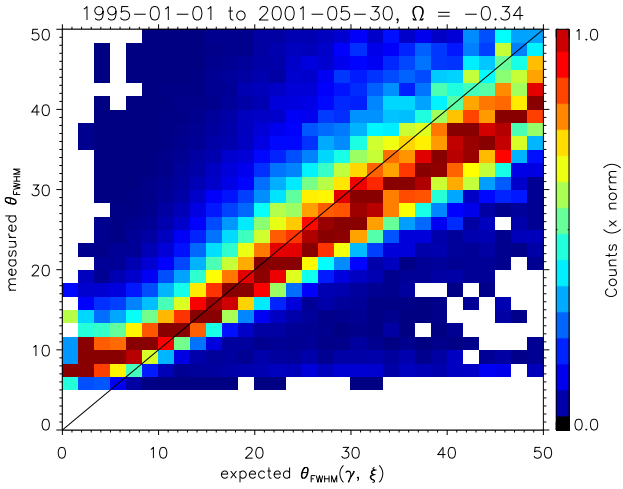}
\caption{\label{strahl_widths_fast_fig} Expected $\theta_{FWHM}$ (degrees) from equation \ref{fwhm_eq} (using $\Omega = -0.34$), plotted versus measured widths. The overplotted line, shown for comparison, represents a one-to-one correspondence.}
\end{figure}

\begin{figure}
\includegraphics[width=1\linewidth]{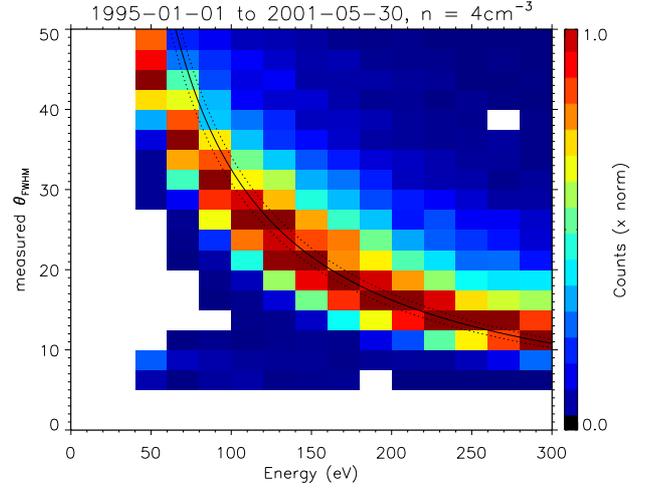}
\caption{\label{strahl_widths_fast_n_fig} Experimental verification of scaling relation (i): $\theta_{FWHM}\propto \mathcal{E}^{-1}$ at fixed $n$. Data shown fall in the range of densities $3.6<n<4.4$cm$^{-3}$.}
\end{figure}

\begin{figure}
\includegraphics[width=1\linewidth]{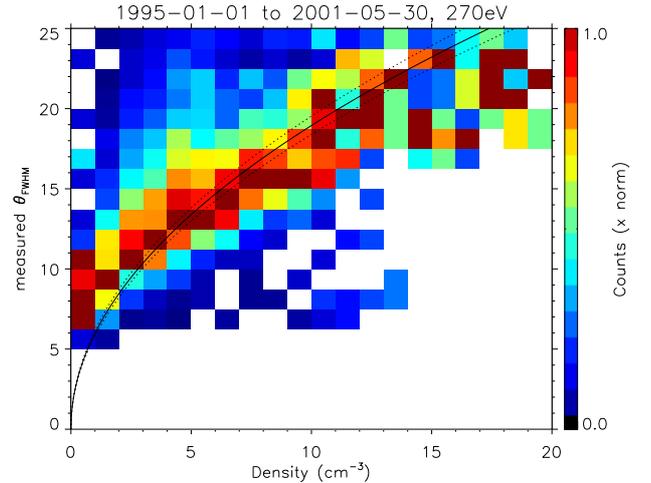}
\caption{\label{strahl_widths_fast_en_fig} Experimental verification of scaling relation (ii): $\theta_{FWHM}\propto \sqrt{n}$ at fixed $\mathcal{E}$. Data shown at detector energy 270 eV (specified by the detector with accuracy $\Delta\mathcal{E}/\mathcal{E}\approx3\%$). }
\end{figure}

In Figure \ref{strahl_widths_fast_fig}, the ``measured $\theta_{FWHM}$'' is compared with the ``expected $\theta_{FWHM}$''. The latter quantity is calculated according to equation \ref{fwhm_eq} from the detector energy $\mathcal{E}$ and plasma electron density $n$, assuming some value for $\Omega$ (which depends on the solar wind's large-scale structure, see eq. \ref{omega_eq}). Figure \ref{strahl_widths_fast_fig} is a normalized 2D histogram, with each column normalized to its peak value.
% The peak locations imply a functional dependence;
The measured $\theta_{FWHM}$ is observed to be proportional to the expected $\theta_{FWHM}$, with a slope near unity under the appropriate choice of the quantity $\Omega$. We here treat the quantity $\Omega$, which appears in equation \ref{fwhm_eq}, as a single constant for all measurements. We set $\Omega = -0.34$ to produce figures \ref{strahl_widths_fast_fig}, \ref{strahl_widths_fast_n_fig}, and \ref{strahl_widths_fast_en_fig}. This value is derived from the average measured $\Omega$ (eq. \ref{omega_m_eq}) among all distributions whose measured widths (eq. \ref{fwhm_m_eq}) fall within a representative range: $8^\circ <\theta_{FWHM} < 50^\circ$.

Our model predicts that $\theta_{FWHM}$ depends on both the electron density $n$ and the detector energy $\mathcal{E}$, through scaling relations (i) and (ii). These dependencies cannot be discriminated in figure \ref{strahl_widths_fast_fig}, so we will now examine them individually. In figure \ref{strahl_widths_fast_n_fig} we demonstrate (i): the strahl width is inversely proportional to the energy, for fixed density. To make this figure, we consider only fast wind data for which the background electron density falls within a narrow range, $3.6<n<4.4$ cm$^{-3}$. For this data, which is effectively a subset of the data shown in figure \ref{strahl_widths_fast_fig}, we plot the measured $\theta_{FWHM}$ versus the detector energy $\mathcal{E}$. The column-normalized 2D histogram nicely matches the predicted trend (equation \ref{fwhm_eq}), which is shown as a solid line for $n=4$ cm$^{-3}$. Similar predictions for $n=3.6,4.4$ cm$^{-3}$ are shown as dotted lines.

We now verify scaling relation (ii), in a similar manner. The SWE strahl detector only sampled the distribution function at discrete energies; in figure \ref{strahl_widths_fast_en_fig}, we only show fast wind data measured at energy $\mathcal{E} = 270$ eV. Here we plot a column-normalized 2D histogram of the measured $\theta_{FWHM}$ versus the electron density $n$. For comparison, we show our model's prediction for the strahl width at this energy, as a solid line. Predictions resulting from varying $\mathcal{E}$ by $\pm 3\%$ (energy range admitted by the detector), are shown as dotted lines.

In figures \ref{strahl_widths_fast_fig}, \ref{strahl_widths_fast_n_fig}, and \ref{strahl_widths_fast_en_fig}, strahl widths are only shown for the regime $\theta_{FWHM}<50^\circ$. This is because the asymptotic formula for the strahl, equation \ref{asymptotic_strahl_eq}, was derived under the assumption $\mu\approx 1$. Including data with pitch angles less than $25^\circ$ corresponds with the regime $\mu > 0.9$, so our assumption is well-satisfied. %test for this when conducting fits?

Our 100,000 fits represent only about 1\% of the distributions ($f_s$) measured by the SWE strahl detector. Despite this low proportion, we believe our data are representative of the fast wind strahl. We note our cleaning procedure (section \ref{strahl_data_sec}) and selection criteria make only a small proportion of the observed $f_s$ suitable for our study. That is, we only consider fast wind strahls that are prominent and resolved by the detector.  This tends to exclude high-energy measurements. For example, SWE/strahl measured $f_s$ at energies $>500$ eV about 38\% of the time; only 5\% of our retained fits are at these energies.
%The number of distributions that can be retained for our study is pared down significantly by our various selection criteria, which are designed to consider only distributions where the strahl is prominent and resolved by the detector.  
%We note that our various selection criteria retain only $\sim1\%$ of the total measurements made by the SWE strahl detector: we consider only fast wind data, in which the strahl is prominent and resolved by the detector, that is well described (as characterized by $\chi^2$) by our model $\ln F(\mu)\propto (1-\mu)$. However, we believe these selection criteria are necessary, and unlikely to bias our description of the strahl in the ambient fast wind.

The data presented in this study include some measurements made during transient events, such as shocks and coronal mass ejections (CMEs). During such events, the eVDF can exhibit properties that are not representative of the ambient fast wind; e.g. ``counterstreaming'' strahls associated with CMEs. Due to the complexity of the task of sorting our data for all events that could potentially exhibit anomalous anti-sunward strahls, we assume such events to be infrequent enough as to not appreciably bias our data. To justify this assumption, we have conducted a preliminary study, in which we repeated the analysis presented in this section, but excluded data measured during times when a CME passed the Earth (as tabulated in \cite{richardsoncane10}). This exclusion had minimal effect on the plots presented in figures \ref{strahl_widths_fast_fig}, \ref{strahl_widths_fast_n_fig}, and \ref{strahl_widths_fast_en_fig}. We note also that of the eVDF measurements used for fitting here, that took place after May 27, 1996 (the date of the first CME in the index), only 16\% were measured during the transit of a near-Earth CME.

\subsection{Fitting to $F_{ave}$}\label{fmodel_fave_sec}

Having analyzed the angular variation of the strahl in detail, we now conduct a new comparison between the data and equation \ref{asymptotic_strahl_eq}, with 2D fits that describe the variation of the eVDF with both angle and energy. We will assume the function $C(v)$ appearing in equation \ref{asymptotic_strahl_eq} is a power law, so that our strahl model can be expressed in the form:
\begin{equation}\label{asymptotic_strahl_eq2}
F(x, \xi, \mu) = C_0 (x/x_0)^{\alpha_s} \xi^{\epsilon}\exp\left\{\tilde{\gamma}(x) \Omega \xi^2 (1-\mu) \right\}.
\end{equation}
Here $\alpha_s$ and $C_0$ are constants to be determined via the fitting procedure, and we introduced the parameter $\epsilon$ for notational convenience:
\begin{equation}\label{epsilon_def}
\epsilon \equiv \alpha^\prime - \alpha + \frac{\alpha_s}{\alpha_T}.
\end{equation}
For convenience, we give here the expressions of $\Omega$, $\epsilon$, and $\tilde{\gamma}(x)$ in terms of the basic parameters $\alpha_n$, $\alpha_T$, $\alpha_B$, and $\alpha_s$: 
\begin{equation}\label{omega_eq}
\Omega = \alpha_B (2\alpha_T - \alpha_n) - \alpha_n - 1,
\end{equation}
where we assumed $\beta =1$, the value for a proton-electron plasma,
\begin{equation}\label{epsilon_eq}
\epsilon = \frac{1}{\alpha_T}\Big(1 + 2\alpha_n - \frac{3}{2}\alpha_T + \alpha_s + \alpha_n \alpha_B - 2 \alpha_T \alpha_B \Big),  %beta = 1
\end{equation}
and
\begin{eqnarray}
\tilde{\gamma}(x) = -\gamma(x) / \alpha_T,
\end{eqnarray}
where the Knudsen number $\gamma(x)$ is defined in Eq.~(\ref{gamma_eq}).

Since each angular distribution $f_s$ was measured at a single energy by the SWE strahl detector, for the purpose of filling out the $\mu$-$\xi$ space with data we must develop a method of combining multiple measurements. We choose to do this by computing an average distribution $F_{ave}(\mu,\xi)$ from the data, which will be used for our 2D fits. $F_{ave}(\mu,\xi)$ is composed from data that are measured in principle at many different times within our $>$6 year  period of study, and may be associated with many different flux tubes. Since we expect the prevalence of the strahl to vary significantly with collisionality, we will only average together distributions that fall within a narrow range of Knudsen numbers.

We first note that the function $F(\bvec{v}/v_{th})$ can be computed from the local distribution $f(\bvec{v})$ by normalizing to the local density and thermal speed \citep{horaites15}:
 
\begin{equation}\label{F_f_eq}
F(\bvec{v}/v_{th}) = \frac{f(\bvec{v}){v_{th}}^3}{n}
\end{equation}   

\noindent Equation \ref{F_f_eq} follows from equation \ref{f_F_eq} and the normalizations $\int f d^3 v = n$, $\int F d^3(v/v_{th}) = 1$. 

\begin{figure*}   % figure spans two columns
\includegraphics[width=1\linewidth]{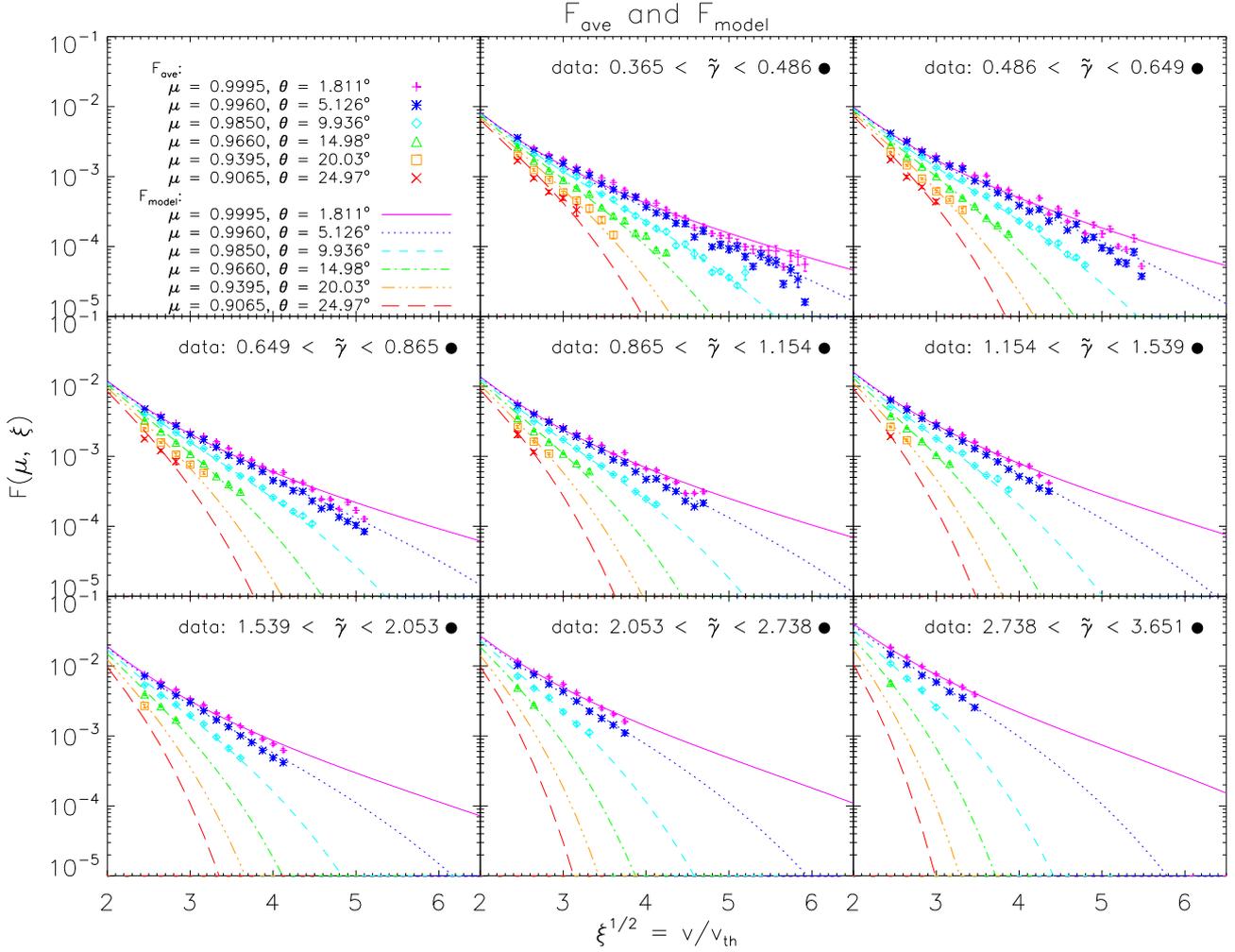}
\caption{\label{fmodel_fave_fig} The 2D strahl distributions $F_{ave}(\mu, \xi)$ are constructed from averaging pitch angle distributions measured by the SWE strahl detector, from fast wind data where $\tilde{\gamma}$ falls within a given range. Data are selected only where the strahl amplitude is sufficiently above the background, and where the strahl is expected to be resolved (see section \ref{fmodel_fave_sec}). Cuts of $F_{ave}$ are shown as points. Fits to a Coulomb scattering model ($F_{model}$), given by equation \ref{asymptotic_strahl_eq2}, are plotted as lines. 
%Fits to a generalized function which captures a variety of scattering theories, ``Model 2'', are shown as dotted lines. %Models 1 and 2 yield similar results, indicating that anomalous scattering is quasi-Coulombic (see section \ref{gen_operator_sec}).
 Parameters of the fits are displayed in table \ref{fit_table}.} 
\end{figure*}

The mean distributions $F_{ave}$ are computed after fitting the angular distributions $f_s$ for the strahl width, a process that was described in section \ref{width_obs_sec}. That is, for each cleaned angular distribution $f_s$, the associated normalized distribution $F_s$ is calculated according to equation \ref{F_f_eq} using the local plasma parameters $n$, $v_{th}$. The angular bins of $F_s$ are assigned coordinates\footnote{The energy $\xi=\mathcal{E}/T$ is calculated from the detector energy $\mathcal{E}$ and local temperature $T$, while $\mu$ depends on each angular bin and on the $\hat B$ direction that was determined during the angular fitting procedure (section \ref{width_obs_sec}).} $\mu$ and $\xi$, which are then sorted into a $\mu$-$\xi$ grid with resolution $\Delta \mu = 0.0005$, $\Delta \xi = 1$. All of the strahl data from our $>$100,000 fits are sorted in this way and averaged by $\mu$-$\xi$ bin to construct the distribution $F_{ave}$. As mentioned, we expect the distribution to depend on the Knudsen number, so we sort the data by $\tilde{\gamma}$ (calculated from $n$ and $T$) when computing $F_{ave}$. We bin by $\tilde{\gamma}$ logarithmically, covering the range $0.365 <\tilde{\gamma} <3.651$ in 8 bins. This range contains $\sim$90\% of our fast wind strahl measurements. Cuts of $F_{ave}$, for these 8 Knudsen numbers, are shown as points in figure \ref{fmodel_fave_fig}. The error bars displayed are the nominal standard deviation of the mean that is computed during the averaging process.
 
Once the average distribution $F_{ave}(\mu,\xi)$ is constructed, we fit the data to our model function, equation \ref{asymptotic_strahl_eq2}.  The parameters $C_0$, $\epsilon$, and $\Omega$ are determined by a nonlinear least squares fit. For the purpose of fitting, $\tilde{\gamma}$ is set to a fixed value, the geometric mean of the maximum and minimum $\tilde{\gamma}$ that were used to bin $F_{ave}$. A comparison between $F_{ave}$ and our fitted model function is shown in figure \ref{fmodel_fave_fig}. In the interest of presenting the data clearly, we choose to fit along only along a few cuts of $F_{ave}$. Namely, we fit to data along the cuts $\mu =$ 0.9995, 0.9960, 0.9850, 0.9660, 0.9395, 0.9065, which corresponds with roughly 5$^\circ$ spacing in pitch angle. These cuts of our fit function are shown as lines in the figure. % in figure \ref{fmodel_fave_fig}.

\begin{table*}
  \centering
%\caption {$F_{ave}$ fit parameters}
\begin{tabular}{| c | c || c | c | c | }
 \hline
 $\tilde{\gamma}$ range ($F_{ave}$) & nominal $\tilde{\gamma}$ ($F_{model}$) & $C_0$ & $\epsilon$ & $\Omega$ \\ % & $\rho$ \\
 \hline
 \hline
\rule{0pt}{4ex}  
 --- & --- & \multicolumn{3}{c |}{model: $F(\mu, \xi) = C_0 \xi^{\epsilon} \exp\{\tilde{\gamma} \Omega \xi^2 (1-\mu) \}$} \\
\rule{0pt}{1ex}  & &  \multicolumn{3}{c |}{} \\
 \hline

$0.365 < \tilde{\gamma} < 0.486$ & $\tilde{\gamma} = 0.421$ & $0.157 \pm 0.011$ & $-2.13 \pm 0.031$ & $-0.38 \pm 0.013$  \\
$0.486 < \tilde{\gamma} < 0.649$ & $\tilde{\gamma} = 0.562$ & $0.191 \pm 0.014$ & $-2.14 \pm 0.033$ & $-0.35 \pm 0.013$  \\
$0.649 < \tilde{\gamma} < 0.865$ & $\tilde{\gamma} = 0.749$ & $0.234 \pm 0.019$ & $-2.14 \pm 0.035$ & $-0.30 \pm 0.012$ \\
$0.865 < \tilde{\gamma} < 1.154$ & $\tilde{\gamma} = 1.000$ & $0.264 \pm 0.020$ & $-2.13 \pm 0.033$ & $-0.28 \pm 0.011$  \\
$1.154 < \tilde{\gamma} < 1.539$ & $\tilde{\gamma} = 1.333$ & $0.306 \pm 0.025$ & $-2.13 \pm 0.036$ & $-0.27 \pm 0.010$  \\
$1.539 < \tilde{\gamma} < 2.053$ & $\tilde{\gamma} = 1.778$ & $0.401 \pm 0.040$ & $-2.19 \pm 0.045$ & $-0.25 \pm 0.011$ \\
$2.053 < \tilde{\gamma} < 2.738$ & $\tilde{\gamma} = 2.371$ & $0.485 \pm 0.053$ & $-2.08 \pm 0.050$ & $-0.28 \pm 0.010$  \\
$2.738 < \tilde{\gamma} < 3.651$ & $\tilde{\gamma} = 3.162$ & $0.669 \pm 0.065$ & $-2.02 \pm 0.046$ & $-0.28 \pm 0.009$  \\

 \hline

 \hline
\end{tabular}
\caption{Model parameters $C_0$, $\epsilon$, $\Omega$ corresponding with fits displayed in figure \ref{fmodel_fave_fig}. The range of $\tilde{\gamma}$ listed in each row represents the Knudsen numbers spanned by the data used to create $F_{ave}$. The column ``nominal $\tilde{\gamma}$'' shows the Knudsen number used for the fit. }
\label{fit_table}
\end{table*}

Some cuts in figure \ref{fmodel_fave_fig} span a larger energy range than others, because the data must satisfy multiple selection criteria to be included in the fit. Two of these selection criteria are based on the expected width of the strahl at each energy $\xi$, which we predicted by assuming $\Omega = -0.34$ (section \ref{width_obs_sec}). First, if $F$ at a given angle is expected to be less than 1/5 the peak ($\mu = 0$) value at that same energy, then the data are not included in the fit; this is to prevent any accidental contamination by the halo.\footnote{see also step ``Remove the halo'' of our data cleaning procedure, section \ref{strahl_data_sec}} Second, if the expected $\theta_{FWHM}$ at a given energy is less than 10$^\circ$, these data are not included; this is to ensure that the measured strahl is resolved by the detector. Additionally, we only include data that falls in the energy range $\xi>5$, because our model equation was derived under the assumption $\xi>>1$.

The results of our fits, for various Knudsen numbers, are summarized in table \ref{fit_table}.
 %in the regime where we are confident that the strahl is resolved by the detector and adequately above the halo background. 
We note that the overall amplitude of the strahl, $C_0$, increases with $\tilde{\gamma}$. This is as expected, since more runaway electrons should be observed as collisionality decreases. The fit parameter $\epsilon$, which dictates the energy dependence of the strahl amplitude, seems fairly independent of $\tilde{\gamma}$, with $\epsilon \approx -2.1$ for all 8 fits. The fit parameter $\Omega$ is also fairly constant, with $\Omega \approx -0.3$ describing most fits well. The fits are in reasonable agreement with our previous estimate $\Omega = -0.34$  (section \ref{width_obs_sec}). 
%We note that these fits to $\Omega$ tend to yield somewhat lower values than $\Omega = -0.34$, which was estimated in section \ref{width_obs_sec}. 

\section{Discussion}\label{discussion_sec}

The shape of the strahl distribution is characterized by the fit parameters $\Omega$ and $\epsilon$ in our model. We would like to determine whether our observed values of $\Omega$ and $\epsilon$ are consistent with the known structure of the solar wind. This structure is described by the parameters $\alpha_n$, $\alpha_T$, and $\alpha_B$. 

The parameters $\Omega$ and $\epsilon$ are measured in this paper, so we will treat them as given. If we select values for two parameters in the set $\{\alpha_n, \alpha_T, \alpha_B, \alpha_s\}$, the remaining two are determined by equations \ref{omega_eq}, \ref{epsilon_eq}. We believe that $\alpha_n$ and $\alpha_B$ are the best-understood of these four, so we will take these values from observations and solve for $\alpha_T$, $\alpha_s$. We can consider our model to be reasonable if all these values are within experimental error.

Inverting equations \ref{omega_eq}, \ref{epsilon_eq}, we find equations for $\alpha_T$ and $\alpha_s$, in terms of the other parameters:
%$\Omega$, $\epsilon$, $\alpha_n$, and $\alpha_B$:

\begin{equation}\label{alpha_t_eq}
%\alpha_T = \frac{\beta \Omega + \alpha_n (\alpha_B + 1) + 1}{2 \alpha_B},
\alpha_T = \frac{\Omega + \alpha_n (\alpha_B + 1) + 1}{2 \alpha_B},       %beta = 1
\end{equation}

\begin{equation}\label{alpha_s_eq}
%\alpha_s = \epsilon \alpha_T + 2 \alpha_T \alpha_B - \alpha_n \alpha_B + \frac{3}{2}\alpha_T - 2 \alpha_n -1.
\alpha_s = \alpha_T(\epsilon + 2\alpha_B + 3/2) - \alpha_n (\alpha_B + 2) -1.
\end{equation}

%\noindent In our model, the parameters $\alpha_n$, $\alpha_T$, $\alpha_B$, and $\alpha_s$ describe the power law variation of $n$, $T$, $B$, and strahl amplitude with distance along a curved flux tube. In practice, however, such power law indices are usually measured with respect to radial variation. 
\noindent Let us see which values of $\alpha_T$, $\alpha_s$ are predicted by our model for the typical fast wind, $\tilde{\gamma}\approx 0.75$. We will at first neglect the flux tube curvature and substitute the observed radial power law indices directly into equations \ref{alpha_t_eq}, \ref{alpha_s_eq}. Let us assume the bulk solar wind is spherically expanding and has reached its asymptotic velocity, so $\alpha_n = -2$ (which follows from the continuity equation). \cite{mariani79} found $B\propto r^{-1.86}$ using Helios 2 fast wind data gathered over the range of heliocentric distances $0.3 AU < r < 1 AU$, so let us assume $\alpha_B = -1.86$. If we substitute these values for $\alpha_n$ and $\alpha_B$ into equation \ref{alpha_t_eq}, while while also letting $\Omega = -0.30$,  $\epsilon=-2.14$ (see table \ref{fit_table}), we predict $\alpha_T = -0.65$. Then it follows $\alpha_s = 2.12$ from equation \ref{alpha_s_eq}.

This predicted value of $\alpha_T$ is consistent with the observed radial scaling of the fast wind core temperature, $T\propto r^{-0.64}$, measured by \cite{pilipp90} using Helios 1 data. These findings are summarized in table \ref{profiles_table}. Here the column ``Measured (radial)'' quotes empirical values, uncertainties, and relevant references. The adjacent column, ``Model (radial)'', presents the set of values described above, that are consistent with our model. %that satisfy equations \ref{alpha_t_eq}, \ref{alpha_s_eq}, described above. 

Technically, the effective power law indices that describe variation along a curved flux tube will differ from the radial indices quoted above. We can estimate these effective indices by assuming the flux tube forms an arm in the Parker spiral. Let us assume that the Parker spiral angle $\theta_p \equiv \hat B \cdot \hat r$ in the ecliptic follows the well-known formula $\tan \theta_p(r) = \omega r / v_{sw}$. This follows from magnetic flux conservation, assuming the nominal angular velocity of the Sun's rotation $\omega$ and the solar wind speed $v_{sw}$ are both constant. Observations show $\theta_P \approx 45^\circ$ at $r=1$ AU \citep[e.g., ][]{luhmann93}, so let us write $\tan \theta_p = r/r_0$, where $r_0 = 1$ AU. We can define the distance $x$ along such a flux tube in terms of $r$, i.e. $x(r) = \int_0^r dr^\prime / \cos \theta_p(r^\prime)$. Evaluating this integral, we find:

\begin{equation}
x(r) = \frac{r_0}{2} \Big\{ \frac{r}{r_0} \sqrt{\Big(\frac{r}{r_0}\Big)^2 + 1}  + \ln \Big[ \frac{r}{r_0} + \sqrt{\Big(\frac{r}{r_0}\Big)^2 + 1} \Big] \Big\}.
\end{equation}

Consider now a physical quantity, $Y$, that varies as a power law with heliocentric distance: $Y(r) = Y_0 (r/r_0)^{\alpha_Y}$, where $Y_0$ and $\alpha_Y$ are constants. The local power law {\it along the flux tube}, $d\ln Y/d\ln x(r)$, evaluated at $r=r_0$ is:

\begin{equation}\label{eq_index_r_to_x}
%\frac{d\ln Y}{d\ln x(r)}\Big|_{r=r_0} = \frac{\sqrt{2} + \ln(1 + \sqrt{2})}{\sqrt{2} + 1 + 1/\sqrt{2}}\alpha_Y \approx 0.735 \alpha_Y
%\frac{d\ln Y}{d\ln x(r)}\Big|_{r=r_0} = \frac{2 + \sqrt{2}\ln(1 + \sqrt{2})}{3 + \sqrt{2}}\alpha_Y \approx 0.735 \alpha_Y.
\frac{d\ln Y}{d\ln x(r)}\Big|_{r=r_0} = \frac{\sqrt{2}+\ln(1 + \sqrt{2})}{2\sqrt{2}}\alpha_Y \approx 0.812 \alpha_Y.
\end{equation}

\noindent We see from equation \ref{eq_index_r_to_x} that a power law with respect to $r$ will appear (locally) as a shallower power law with respect to $x$. In the column ``Measured (curvilinear)'' of table \ref{profiles_table}, we correct for the curvature of the Parker spiral at 1 AU by multiplying the values of $\alpha_n$, $\alpha_T$ and $\alpha_B$ reported in column ``Measured (radial)'' by a factor of 0.812. This describes the effective power law that is relevant to our solutions to the drift-kinetic equation. We then repeat our previous analysis, by setting $\alpha_n$, $\alpha_B$, $\Omega$, $\epsilon$ to their measured values, and solving for $\alpha_T$ and $\alpha_s$ from equations \ref{alpha_t_eq}, \ref{alpha_s_eq}. The resulting values, shown in column ``Model (curvilinear)'' of table \ref{profiles_table}, are self-consistent with regards to our model and are in good agreement with observations of the solar wind's large-scale structure.

Table \ref{profiles_table} serves to illustrate the ease with which observations of the strahl shape can be explained in terms the solar wind's large scale structure. We note that the values of these structural parameters are subject to appreciable experimental uncertainty, and the ``true'' values may be significantly different from those quoted here. In particular, different published measurements of $\alpha_T$ in the fast wind near 1 AU vary quite widely, with some measurements falling well outside the error bars quoted from our reference, \cite{pilipp90}. For a review of published measurements of $\alpha_T$, see \cite{maksimovic00}. We note it is generally agreed that temperature varies with a significantly flatter profile than $\alpha_T = -4/3$, the value associated with adiabatic expansion.

\begin{table*}
  \centering
%\caption {$F_{ave}$ fit parameters}
\begin{tabular}{| c || c | c || c | c | }
%\begin{tabular}{| c | c || c | c | c || c | c | c | c | }
 \hline
 --- & Measured (radial) & Model (radial) & Measured (curvilinear) & Model (curvilinear) \\
 \hline

%$\Omega$ & $-0.27 \pm 0.01$ (present paper)                         & -0.27  & $ -0.27  \pm 0.01$ (present paper) & -0.27 \\
%$\epsilon$ & $-2.21 \pm 0.03$ (present paper)                      & -2.21 & $ -2.21 \pm 0.03 $ (present paper) & -2.21 \\
$\Omega$ & $-0.30 \pm 0.01$ (present paper)                         & -0.30  & $ -0.30  \pm 0.01$ (present paper) & -0.30 \\
$\epsilon$ & $-2.14 \pm 0.04$ (present paper)                      & -2.14 & $ -2.14 \pm 0.04 $ (present paper) & -2.14 \\
$\alpha_n$ & -2                                               & -2 & $ -1.62 $ & -1.62\\
%note pilipp 1990 assumed 10% error, see pg. 6322
%$\alpha_B$ & $-1.58 \pm 0.10$ (Mariani 1979)              & -1.58 & $ -1.16 \pm 0.07 $ & -1.16\\
$\alpha_B$  & $-1.86 \pm 0.05$ \citep{mariani79}              &  -1.86     & $-1.51 \pm 0.04$ & -1.51\\
 \hline
%$\alpha_T$ & $-0.64 \pm 0.06$    \citep{pilipp90}           & -0.66 & $ -0.47 \pm 0.04 $ & -0.46\\
$\alpha_T$ & $-0.64 \pm 0.06$    \citep{pilipp90}           & -0.65 & $ -0.52 \pm 0.05 $ & -0.51\\
%           & $-0.48 \pm 0.05$     (H2, Pilipp 1990)           &       & $ \pm $ & \\
%$\alpha_s$ & see ``Discussion''                         & \bf{2.20}  &  see ``Discussion'' & \bf{1.53}\\
$\alpha_s$ & see ``Discussion''                         & \bf{2.12}  &  see ``Discussion'' & \bf{1.65}\\
 \hline
\end{tabular}
\caption{Comparison between observations of the large-scale structure in the fast solar wind and a set of parameters that is consistent with our model (eq. \ref{asymptotic_strahl_eq2}). The relevant reference for each nominal observational value is given in parentheses in the ``Measured (radial)'' column. Values of $\alpha_n$, $\alpha_T$, $\alpha_B$ in the ``Measured (curvilinear)'' column are found by multiplying the corresponding values in the ``Measured (radial)'' column by 0.812 (eq. \ref{eq_index_r_to_x}). This approximates the local power law along a curved Parker spiral flux tube. Each ``Model'' column is identical to the corresponding ``Measured'' column, except for the values of $\alpha_T$ and $\alpha_s$, which are obtained from equations \ref{alpha_t_eq}, \ref{alpha_s_eq}, given the other model parameters.}
\label{profiles_table}
\end{table*}

In both the radial and curvilinear models displayed in Table \ref{profiles_table}, we predict $\alpha_s > 0$. The positive sign of $\alpha_s$ means that the amplitude of the strahl (in our normalized distribution $F$) should increase with heliocentric distance $r$. This prediction holds even if we allow the input parameters of our model to vary by $\pm 20\%$. We highlight this prediction because it is in some conflict with the observations of \cite{maksimovic05} and \cite{stverak09}, which showed that the strahl density (relative to core density) actually {\it decreases} with $r$.

To address this point, let us consider Figure 10 of \cite{stverak09}. The upper right panel of that figure shows the parallel cuts of the average fast wind eVDF at different heliocentric distances. The eVDF can be roughly interpreted as a cut of $F(\bvec{v}/v_{th})$, up to an amplitude coefficient, since the velocities shown on the x-axis are normalized to the core thermal speed, as in our equation \ref{f_F_eq}. Their figure indeed shows that the strahl diminishes with $r$, but only in the velocity range $v_{\parallel}/v_{th} \lesssim 5$. We note that the trend is fairly weak, so that although $\alpha_s$ is negative we can estimate $|\alpha_s| < 1$ from their plot. At higher velocities the opposite trend can be observed: the relative strahl amplitude actually increases with distance, in agreement with our model. We note that this regime $v \gg v_{th}$ is actually where our model is most applicable, since in our derivation we assumed $\xi \gg 1$ and dropped terms that were low order in $\xi$ (e.g. the electric field terms).  The strahl is certainly present at these energies\footnote{ See e.g., Figure 4, \cite{ogilvie00}. On May 11, 1999, a narrow strahl beam was observed up to the highest energy sampled by the SWE strahl detector, $\sim$1200 eV. Comparing with the local temperature $T$, this energy corresponds with $v/v_{th}\approx 6.5$. We have checked this distribution by hand (not shown), which indeed exhibits a high-energy strahl. However, we believe the distribution is too narrow to be resolved by the detector. Indeed, if the strahl continues to narrow at high energies, the detector may not register significant counts even if the field-parallel (peak) amplitude of the strahl is significant, and falsely appear to be subsumed by the halo.}, and is well above the halo background. However, only some of our current measurements apply to this regime; velocities only as high as $v\approx 6 v_{th}$ are used in our fits to $F_{ave}$ (figure \ref{fmodel_fave_fig}). As discussed in section \ref{fmodel_fave_sec}, these data are excluded because the strahl becomes unresolved at high energies. We conclude that the rate of strahl growth relative to the core ($\sim$$\alpha_s$), at both intermediate and very high strahl energies, deserves further empirical investigation.

Since current empirical evidence indicates that the relative amplitude of the strahl at intermediate energies diminishes with inceasing $r$, we must account for this discrepancy between the data and our model. One possible explanation is that the regime $v \lesssim 5 v_{th}$ is not truly asymptotic, so that the electric field terms (which we neglected) should be considered at these energies. Unfortunately, if we include the electric field terms in the steps following equation \ref{ssk_eq} in section \ref{theory_sec}, we run into two difficulties. First, the spatial dependence of the ambipolar field $\gamma_E(x)$ must be assumed, and this quantity is model-dependent and not directly measureable by current satellites. Second, there is no guarantee that closed form analytic solutions, analogous to equation \ref{asymptotic_strahl_exact_eq}, can be found if the electric field terms are included. The problem appears intractable, for instance, if we assume $\gamma_E \propto \gamma$. We note ambipolar electric forces can be included in kinetic simulations \citep[e.g., ][]{smith12}, so that the effect of $\gamma_E$ may be understood through numerical analysis.

Finally, we note that other diffusion mechanisms, such as wave-particle interactions, may affect the solar wind strahl. Indeed, \cite{stverak09} suggest the diminishment of the relative strahl density with distance, and the corresponding increase in halo density, might indicate that wave-particle interactions scatter strahl electrons into the halo population \citep[see also][]{gurgiolo12}.  We do not make an effort to include wave-particle scattering into our current model, as this would greatly complicate our discussion. However, we do note that if such interactions can be expressed in terms of a pitch angle diffusion operator, it may be possible to predict a form for the strahl distribution via methods analogous to those introduced in section \ref{theory_sec}.

\section{Summary and Conclusions}

The theory developed in this paper should be understood as a reduced model of solar wind electron kinetics. Our model assumes an ordered structure of the large-scale profiles of $n$, $T$, and $B$, which allows the drift kinetic equation to be written in a tractable form. This yields, for the first time, a simple analytic prediction for the shape of the strahl distribution (equation \ref{asymptotic_strahl_eq}). In the present paper we show this prediction captures the essential features of strahl eVDFs in the ambient fast wind. In particular, we find that at a given energy, the strahl is approximately Gaussian with respect to the pitch angle $\theta$. We obtain scaling relations (i) and (ii), which describe how the beam width $\theta_{FWHM}$ varies with energy and density. We verify the accuracy of our model by comparing it with data collected by the Wind satellite's SWE strahl detector, which sampled the eVDF at high angular resolution. 

%Our kinetic model describes the diffusion of the strahl due to Coulomb collisions, and its
Our model's success in predicting the strahl widths strongly supports the view that Coulomb collisions are an important source of pitch angle scattering.  We note also that the temperature Knudsen number $\gamma$, which describes Coulomb collisionality, plays a central role in our model. Indeed, $\gamma$ appears to be an important parameter for ordering strahl eVDFs (figure \ref{fmodel_fave_fig}), and we observe the strahl amplitude ($C_0$) is correlated with $\gamma$ (table \ref{fit_table}). However, we cannot say definitively that strahl scattering is only due to Coulomb collisions, as it seems our model's predictiveness might be improved by incorporating another physical effect. % Like \cite{lemonsfeldman83}, we consider wave-particle interactions to be a potentially important source of strahl scattering.

Our observations show that the strahl narrows with energy, in agreement with previous measurements \citep[e.g., ][]{feldman78, pilipp87, fitzenreiter98, ogilvie00}. However, one survey of strahl eVDFs measured at 1 AU by ACE \citep{anderson12} indicated that the strahl width broadens 40$\%$ and narrows 60$\%$ of the time. This trend is not seen in our current observations of the fast wind, in which the strahl appears systematically narrower at higher energies. We have conducted a preliminary investigation of SWE strahl detector data measured in the slow wind (not shown), and again find narrower strahls at higher energies. The discrepancy with \citep{anderson12} may be due to differences in methodology and data selection---their study presented a broad survey of strahl data, whereas the current work deals with narrow, prominent strahls in the fast wind. We note that the SWE strahl detector samples only a $\sim$50$\times$60 degree field of view, so that we may have difficulty identifying very broad strahls. Finally, we note that identification of the strahl, as distinguished from the core and halo populations that also help form the distribution, is in part a matter of definition \citep[see e.g., ][]{stverak09}.

%Additionally, we note that the distinction between the halo and strahl is somewhat arbitrary, especially since a theor

 The presence of the arbitrary function $C(v)$ makes our model (eq. \ref{asymptotic_strahl_eq}) quite flexible, in that it can be matched to any energy spectrum observed at 1 AU. The function $C(v)$  carries information about the electrons' origin---as argued in e.g., \cite{smith12}, the energy spectrum of the strahl may be a vestige of the thermal distribution of electrons near the base of the corona. In practice, we find that $C(v)$ can be matched well to a power law (eq. \ref{asymptotic_strahl_eq2}). However, matching our model to the eVDFs at 1 AU leads us to predict that the relative strahl amplitude should grow with heliocentric distance ($\alpha_s>0$), which appears to conflict with observations. So, our model may require further theoretical development, motivated by detailed observations of the strahl over a range of heliocentric distances. Like \cite{lemonsfeldman83}, we consider wave-particle interactions to be a potentially important source of strahl scattering.

As discussed in \cite{saitogary07}, the energy dependence of strahl widths is intimately connected with the energy dependence of the diffusion operator. For diffusion caused by turbulent wave-particle interactions, the relevant diffusion operator in the kinetic equation is determined by the properties of the waves and their turbulent spectrum. Therefore, the observed shape of the strahl should help direct the search for theories of anomalous diffusion caused by wave-particle scattering. We intend to discuss this in future work.

\appendix

\section{}\label{appendix}

As mentioned in section \ref{theory_sec}, we may reduce equation \ref{asymptotic_strahl_exact_eq} to a scale-invariant form by assuming either $\beta C_2(v)\xi^{-\alpha^\prime}/G(v) \gg 1$ or $\beta C_2(v)\xi^{-\alpha^\prime}/G(v) \ll 1$. Our model function (\ref{asymptotic_strahl_eq}) corresponds with the former case. We will now consider the latter case, and show that it is inconsistent with our observations of the solar wind strahl.

Let us assume 

\begin{equation}\label{assumption_case2}
\beta C_2(v)\xi^{-\alpha^\prime}/G(v) \ll 1,
\end{equation}

\noindent so that equation \ref{asymptotic_strahl_exact_eq} can be written approximately as

\begin{equation}\label{asymptotic_strahl_eq_case2}
F(v, \xi, \mu) = C_1(v)\xi^{- \alpha}\exp\left\{\alpha^{\prime} C_2(v) \xi^{\kappa-\alpha^{\prime}} (1-\mu) \right\},
\end{equation}

\noindent where $C_1(v)$ and $C_2(v)$ are arbitrary functions. If we try to match equation \ref{asymptotic_strahl_eq_case2} to the observed strahl, the function $C_2(v)$ can be constrained by recalling scaling relation (i): the strahl width is inversely proportional to the energy (figure \ref{strahl_widths_fast_n_fig}). If we recast equation \ref{asymptotic_strahl_eq_case2} in terms of $x$, $\xi$, $\mu$, we then require an overall factor $\xi^2$ to appear inside of the exponential function. This can only be accomplished if we assume $C_2(v)$ has the form: %$C_2(v) \propto G(v) v^{2 \alpha^{\prime}} $ has the form: 
%, where the function $G(v)$ is given by equation (\ref{g_v_eq}).

\begin{equation}\label{c2_form_eq}
\begin{split}
C_2(v) &= C_3 G(v) \Big(\frac{v}{v_{th,0}}\Big)^{2 \alpha^\prime}\\
       &= C_3 G(v) \xi^{\alpha^\prime} \Big(\frac{x}{x_0}\Big)^{\alpha^\prime \alpha_T}.
\end{split}
\end{equation}

\noindent Here, $C_3$ is a dimensionless constant, and the function $G(v)$ is given by equation \ref{g_v_eq}. Substituting (\ref{c2_form_eq}) into equation \ref{asymptotic_strahl_eq_case2} yields a prediction for the strahl width. At the position of the Wind satellite, $x=x_0$, this formula for the angular variation of the strahl at a given energy $\xi$ can be written: 

\begin{equation}\label{f_mu_eq_case2}
\begin{split}
%F(x, \xi, \mu) = C(x,\xi) \exp\{C_3 \alpha^\prime \gamma(x) \Big(\frac{x}{x_0}\Big)^{\alpha^\prime \alpha_T} \xi^2 (1-\mu) \} \xi^{-\alpha},
F(\mu) &\propto \exp\{C_3 \alpha^\prime \gamma(x_0) \xi^2 (1-\mu) \}.\\
       &=\exp\{ \tilde{\gamma}(x_0) \Omega^\prime \xi^2 (1-\mu) \}
\end{split}
\end{equation}

Here we introduced the notation $\Omega^\prime \equiv -C_3 \alpha^\prime \alpha_T$. Equation \ref{f_mu_eq_case2} is analogous to equation \ref{f_mu_eq}, and has a similar form. If we were to fit the strahl distribution $f_s$ to (\ref{f_mu_eq_case2}), we would find values of $\Omega^\prime$ identical to the values of $\Omega$ found in section \ref{width_obs_sec}. So let us identify $\Omega \sim \Omega^\prime$, and estimate the value of $C_3$ that would be consistent with our measurements:
%we would find $\Omega^\prime \approx -0.3$, as in section \ref{width_obs_sec}. Referring to table \ref{profiles_table}, we can estimate $\alpha^\prime=-0.58$ from the values $\alpha_n$, $\alpha_T$, and $\alpha_B$ presented in the ``Model (curvilinear)'', which are consistent with satellite observations (section \ref{discussion_sec}). Assuming $\alpha_T = 0.3$ in the typical fast wind, we can estimate the value of $C_3$:

\begin{equation}
C_3 = \frac{-\Omega^\prime}{\alpha^\prime \alpha_T} \sim 1/\beta
\end{equation}

%This contradicts our assumption $C_3 \ll 1/\beta$ (recall $\beta \approx 1$), which is tantamount to contradicting (\ref{assumption_case2}). 
However, from (\ref{c2_form_eq}), the scale-invariant assumption (\ref{assumption_case2}) reduces to the requirement $C_3 \ll 1/\beta$, at the position of the Wind satellite $x=x_0$. We therefore have a contradiction, and judge eq. \ref{asymptotic_strahl_eq_case2} to be less consistent with observed solar wind profiles than eq. \ref{asymptotic_strahl_eq}. In other words: the measured solar wind profiles ($\alpha_n$, $\alpha_T$, $\alpha_B$) and strahl widths (characterized by $\Omega$, see eq. \ref{omega_eq}) conform well with the model studied in this paper (see table \ref{profiles_table})---these same parameters could not be explained by a scale-invariant solution based on the assumption (\ref{assumption_case2}).
%In other words: the measured solar wind profiles ($\alpha_n$, $\alpha_T$, $\alpha_B$) predict strahl widths (characterized by $\Omega$, see eq. \ref{omega_eq}) that conform well with the model studied in this paper (see table \ref{profiles_table})---these same profiles could not explain the data in a scale-invariant solution based on the assumption (\ref{assumption_case2}).

%\begin{acknowledgments}
{\em Acknowledgments}---The authors thank numerous people associated with the Wind spacecraft at the Goddard Space Flight Center--Richard Fitzenreiter, Adam Szabo, Robert Candey, Howard Leckner, and Matt Holland--for their assistance in accessing and interpreting data from the SWE strahl detector. This work was partly supported by the National Science Foundation under the grant NSF PHY-1707272. SB was partly supported by the National Science Foundation under the grant NSF AGS-1261659 and by the Vilas Associates Award from the University of Wisconsin - Madison.
%{\em Acknowledgments}---The authors thank numerous people associated with the Wind spacecraft at the Goddard Space Flight Center--Lynn Wilson, Richard Fitzenreiter, Jan Merka, Adolfo Vi\~nas, Adam Szabo, Robert Candey, Howard Leckner, and Matt Holland--for their assistance in accessing and interpreting data from the SWE strahl detector. SB was partly supported by the National Science Foundation under the grant NSF AGS-1261659 and by the Vilas Associates Award from the University of Wisconsin - Madison.

%\end{acknowledgments}


%\bibliographystyle{mnras}
%\bibliography{strahl_paper_refs} % if your bibtex file is called example.bib

\begin{thebibliography}{}
\makeatletter
\relax
\def\mn@urlcharsother{\let\do\@makeother \do\$\do\&\do\#\do\^\do\_\do\%\do\~}
\def\mn@doi{\begingroup\mn@urlcharsother \@ifnextchar [ {\mn@doi@}
  {\mn@doi@[]}}
\def\mn@doi@[#1]#2{\def\@tempa{#1}\ifx\@tempa\@empty \href
  {http://dx.doi.org/#2} {doi:#2}\else \href {http://dx.doi.org/#2} {#1}\fi
  \endgroup}
\def\mn@eprint#1#2{\mn@eprint@#1:#2::\@nil}
\def\mn@eprint@arXiv#1{\href {http://arxiv.org/abs/#1} {{\tt arXiv:#1}}}
\def\mn@eprint@dblp#1{\href {http://dblp.uni-trier.de/rec/bibtex/#1.xml}
  {dblp:#1}}
\def\mn@eprint@#1:#2:#3:#4\@nil{\def\@tempa {#1}\def\@tempb {#2}\def\@tempc
  {#3}\ifx \@tempc \@empty \let \@tempc \@tempb \let \@tempb \@tempa \fi \ifx
  \@tempb \@empty \def\@tempb {arXiv}\fi \@ifundefined
  {mn@eprint@\@tempb}{\@tempb:\@tempc}{\expandafter \expandafter \csname
  mn@eprint@\@tempb\endcsname \expandafter{\@tempc}}}

\bibitem[\protect\citeauthoryear{{Anderson}, {Skoug}, {Steinberg}  \&
  {McComas}}{{Anderson} et~al.}{2012}]{anderson12}
{Anderson} B.~R.,  {Skoug} R.~M.,  {Steinberg} J.~T.,   {McComas} D.~J.,  2012,
  \mn@doi [Journal of Geophysical Research (Space Physics)]
  {10.1029/2011JA017269}, \href
  {http://adsabs.harvard.edu/abs/2012JGRA..117.4107A} {117, A04107}

\bibitem[\protect\citeauthoryear{{Bale}, {Pulupa}, {Salem}, {Chen}  \&
  {Quataert}}{{Bale} et~al.}{2013}]{bale13}
{Bale} S.~D.,  {Pulupa} M.,  {Salem} C.,  {Chen} C.~H.~K.,   {Quataert} E.,
  2013, \mn@doi [\apjl] {10.1088/2041-8205/769/2/L22}, \href
  {http://adsabs.harvard.edu/abs/2013ApJ...769L..22B} {769, L22}

\bibitem[\protect\citeauthoryear{{Boldyrev}, {Horaites}, {Xia}  \&
  {Perez}}{{Boldyrev} et~al.}{2013}]{boldyrev13}
{Boldyrev} S.,  {Horaites} K.,  {Xia} Q.,   {Perez} J.~C.,  2013, \mn@doi
  [\apj] {10.1088/0004-637X/777/1/41}, \href
  {http://adsabs.harvard.edu/abs/2013ApJ...777...41B} {777, 41}

\bibitem[\protect\citeauthoryear{{Cowie} \& {McKee}}{{Cowie} \&
  {McKee}}{1977}]{cowiemckee77}
{Cowie} L.~L.,  {McKee} C.~F.,  1977, \mn@doi [\apj] {10.1086/154911}, \href
  {http://adsabs.harvard.edu/abs/1977ApJ...211..135C} {211, 135}

\bibitem[\protect\citeauthoryear{{DeForest}}{{DeForest}}{1972}]{deforest72}
{DeForest} S.~E.,  1972, \mn@doi [\jgr] {10.1029/JA077i004p00651}, \href
  {http://adsabs.harvard.edu/abs/1972JGR....77..651D} {77, 651}

\bibitem[\protect\citeauthoryear{{Fairfield} \& {Scudder}}{{Fairfield} \&
  {Scudder}}{1985}]{fairfieldscudder85}
{Fairfield} D.~H.,  {Scudder} J.~D.,  1985, \mn@doi [\jgr]
  {10.1029/JA090iA05p04055}, \href
  {http://adsabs.harvard.edu/abs/1985JGR....90.4055F} {90, 4055}

\bibitem[\protect\citeauthoryear{{Feldman}, {Asbridge}, {Bame}, {Gosling}  \&
  {Lemons}}{{Feldman} et~al.}{1978}]{feldman78}
{Feldman} W.~C.,  {Asbridge} J.~R.,  {Bame} S.~J.,  {Gosling} J.~T.,   {Lemons}
  D.~S.,  1978, \mn@doi [\jgr] {10.1029/JA083iA11p05285}, \href
  {http://adsabs.harvard.edu/abs/1978JGR....83.5285F} {83, 5285}

\bibitem[\protect\citeauthoryear{{Fitzenreiter}, {Ogilvie}, {Chornay}  \&
  {Keller}}{{Fitzenreiter} et~al.}{1998}]{fitzenreiter98}
{Fitzenreiter} R.~J.,  {Ogilvie} K.~W.,  {Chornay} D.~J.,   {Keller} J.,  1998,
  \mn@doi [\grl] {10.1029/97GL03703}, \href
  {http://adsabs.harvard.edu/abs/1998GeoRL..25..249F} {25, 249}

\bibitem[\protect\citeauthoryear{{Gary}, {Feldman}, {Forslund}  \&
  {Montgomery}}{{Gary} et~al.}{1975}]{gary75}
{Gary} S.~P.,  {Feldman} W.~C.,  {Forslund} D.~W.,   {Montgomery} M.~D.,  1975,
  \mn@doi [\jgr] {10.1029/JA080i031p04197}, \href
  {http://adsabs.harvard.edu/abs/1975JGR....80.4197G} {80, 4197}

\bibitem[\protect\citeauthoryear{{Gary}, {Scime}, {Phillips}  \&
  {Feldman}}{{Gary} et~al.}{1994}]{gary94}
{Gary} S.~P.,  {Scime} E.~E.,  {Phillips} J.~L.,   {Feldman} W.~C.,  1994,
  \mn@doi [\jgr] {10.1029/94JA02067}, \href
  {http://adsabs.harvard.edu/abs/1994JGR....9923391G} {99, 23391}

\bibitem[\protect\citeauthoryear{{Gosling}, {Thomsen}, {Bame}  \&
  {Zwickl}}{{Gosling} et~al.}{1987}]{gosling87}
{Gosling} J.~T.,  {Thomsen} M.~F.,  {Bame} S.~J.,   {Zwickl} R.~D.,  1987,
  \mn@doi [\jgr] {10.1029/JA092iA11p12399}, \href
  {http://adsabs.harvard.edu/abs/1987JGR....9212399G} {92, 12399}

\bibitem[\protect\citeauthoryear{{Graham} et~al.,}{{Graham}
  et~al.}{2017}]{graham17}
{Graham} G.~A.,  et~al., 2017, \mn@doi [Journal of Geophysical Research (Space
  Physics)] {10.1002/2016JA023656}, \href
  {http://adsabs.harvard.edu/abs/2017JGRA..122.3858G} {122, 3858}

\bibitem[\protect\citeauthoryear{{Gurevich} \& {Istomin}}{{Gurevich} \&
  {Istomin}}{1979}]{gurevichistomin79}
{Gurevich} A.~V.,  {Istomin} Y.~N.,  1979, Soviet Journal of Experimental and
  Theoretical Physics, \href
  {http://adsabs.harvard.edu/abs/1979JETP...50..470G} {50, 470}

\bibitem[\protect\citeauthoryear{{Gurgiolo}, {Goldstein}, {Vi{\~n}as}  \&
  {Fazakerley}}{{Gurgiolo} et~al.}{2012}]{gurgiolo12}
{Gurgiolo} C.,  {Goldstein} M.~L.,  {Vi{\~n}as} A.~F.,   {Fazakerley} A.~N.,
  2012, \mn@doi [Annales Geophysicae] {10.5194/angeo-30-163-2012}, \href
  {http://adsabs.harvard.edu/abs/2012AnGeo..30..163G} {30, 163}

\bibitem[\protect\citeauthoryear{{Hammond}, {Feldman}, {McComas}, {Phillips}
  \& {Forsyth}}{{Hammond} et~al.}{1996}]{hammond96}
{Hammond} C.~M.,  {Feldman} W.~C.,  {McComas} D.~J.,  {Phillips} J.~L.,
  {Forsyth} R.~J.,  1996, \aap, \href
  {http://adsabs.harvard.edu/abs/1996A%26A...316..350H} {316, 350}

\bibitem[\protect\citeauthoryear{{Helander} \& {Sigmar}}{{Helander} \&
  {Sigmar}}{2002}]{helandersigmar02}
{Helander} P.,  {Sigmar} D.~J.,  2002, {Collisional transport in magnetized
  plasmas}

\bibitem[\protect\citeauthoryear{{Hollweg}}{{Hollweg}}{1970}]{hollweg70}
{Hollweg} J.~V.,  1970, \mn@doi [\jgr] {10.1029/JA075i013p02403}, \href
  {http://adsabs.harvard.edu/abs/1970JGR....75.2403H} {75, 2403}

\bibitem[\protect\citeauthoryear{{Horaites}, {Boldyrev}, {Krasheninnikov},
  {Salem}, {Bale}  \& {Pulupa}}{{Horaites} et~al.}{2015}]{horaites15}
{Horaites} K.,  {Boldyrev} S.,  {Krasheninnikov} S.~I.,  {Salem} C.,  {Bale}
  S.~D.,   {Pulupa} M.,  2015, \mn@doi [Physical Review Letters]
  {10.1103/PhysRevLett.114.245003}, \href
  {http://adsabs.harvard.edu/abs/2015PhRvL.114x5003H} {114, 245003}

\bibitem[\protect\citeauthoryear{{Jockers}}{{Jockers}}{1970}]{jockers70}
{Jockers} K.,  1970, \aap, \href
  {http://adsabs.harvard.edu/abs/1970A%26A.....6..219J} {6, 219}

\bibitem[\protect\citeauthoryear{{Kajdi{\v c}}, {Alexandrova}, {Maksimovic},
  {Lacombe}  \& {Fazakerley}}{{Kajdi{\v c}} et~al.}{2016}]{kajdic16}
{Kajdi{\v c}} P.,  {Alexandrova} O.,  {Maksimovic} M.,  {Lacombe} C.,
  {Fazakerley} A.~N.,  2016, \mn@doi [\apj] {10.3847/1538-4357/833/2/172},
  \href {http://adsabs.harvard.edu/abs/2016ApJ...833..172K} {833, 172}

\bibitem[\protect\citeauthoryear{{Krasheninnikov}}{{Krasheninnikov}}{1988}]{krasheninnikov88}
{Krasheninnikov} S.~I.,  1988, \mn@doi [JETP] {10.1103/PhysRev.89.977}, \href
  {http://adsabs.harvard.edu/abs/1953PhRv...89..977S} {67, 2483}

\bibitem[\protect\citeauthoryear{{Krasheninnikov} \&
  {Bakunin}}{{Krasheninnikov} \& {Bakunin}}{1993}]{krasheninnikovbakunin93}
{Krasheninnikov} S.~I.,  {Bakunin} O.~G.,  1993, Technical report,
  {Self-similar variables and the problem of nonlocal electron heat
  conductivity (MIT Plasma Fusion Center)}

\bibitem[\protect\citeauthoryear{{Kulsrud}}{{Kulsrud}}{1983}]{kulsrud83}
{Kulsrud} R.~M.,  1983, in {Galeev} A.~A.,  {Sudan} R.~N.,  eds, Basic Plasma
  Physics: Selected Chapters, Handbook of Plasma Physics, Volume 1 (North
  Holland Publishing Company, Amsterdam). p.~1

\bibitem[\protect\citeauthoryear{{Lacombe}, {Alexandrova}, {Matteini},
  {Santol{\'{\i}}k}, {Cornilleau-Wehrlin}, {Mangeney}, {de Conchy}  \&
  {Maksimovic}}{{Lacombe} et~al.}{2014}]{lacombe14}
{Lacombe} C.,  {Alexandrova} O.,  {Matteini} L.,  {Santol{\'{\i}}k} O.,
  {Cornilleau-Wehrlin} N.,  {Mangeney} A.,  {de Conchy} Y.,   {Maksimovic} M.,
  2014, \mn@doi [\apj] {10.1088/0004-637X/796/1/5}, \href
  {http://adsabs.harvard.edu/abs/2014ApJ...796....5L} {796, 5}

\bibitem[\protect\citeauthoryear{{Lemons} \& {Feldman}}{{Lemons} \&
  {Feldman}}{1983}]{lemonsfeldman83}
{Lemons} D.~S.,  {Feldman} W.~C.,  1983, \mn@doi [\jgr]
  {10.1029/JA088iA09p06881}, \href
  {http://adsabs.harvard.edu/abs/1983JGR....88.6881L} {88, 6881}

\bibitem[\protect\citeauthoryear{{Lepping} et~al.,}{{Lepping}
  et~al.}{1995}]{lepping95}
{Lepping} R.~P.,  et~al., 1995, \mn@doi [\ssr] {10.1007/BF00751330}, \href
  {http://adsabs.harvard.edu/abs/1995SSRv...71..207L} {71, 207}

\bibitem[\protect\citeauthoryear{{Luhmann}, {Zhang}, {Petrinec}, {Russell},
  {Gazis}  \& {Barnes}}{{Luhmann} et~al.}{1993}]{luhmann93}
{Luhmann} J.~G.,  {Zhang} T.-L.,  {Petrinec} S.~M.,  {Russell} C.~T.,  {Gazis}
  P.,   {Barnes} A.,  1993, \mn@doi [\jgr] {10.1029/92JA02235}, \href
  {http://adsabs.harvard.edu/abs/1993JGR....98.5559L} {98, 5559}

\bibitem[\protect\citeauthoryear{{Maksimovic}, {Gary}  \& {Skoug}}{{Maksimovic}
  et~al.}{2000}]{maksimovic00}
{Maksimovic} M.,  {Gary} S.~P.,   {Skoug} R.~M.,  2000, \mn@doi [\jgr]
  {10.1029/2000JA900039}, \href
  {http://adsabs.harvard.edu/abs/2000JGR...10518337M} {105, 18337}

\bibitem[\protect\citeauthoryear{{Maksimovic} et~al.,}{{Maksimovic}
  et~al.}{2005}]{maksimovic05}
{Maksimovic} M.,  et~al., 2005, \mn@doi [Journal of Geophysical Research (Space
  Physics)] {10.1029/2005JA011119}, \href
  {http://adsabs.harvard.edu/abs/2005JGRA..11009104M} {110, 9104}

\bibitem[\protect\citeauthoryear{{Mariani}, {Villante}, {Bruno}, {Bavassano}
  \& {Ness}}{{Mariani} et~al.}{1979}]{mariani79}
{Mariani} F.,  {Villante} U.,  {Bruno} R.,  {Bavassano} B.,   {Ness} N.~F.,
  1979, \mn@doi [\solphys] {10.1007/BF00174545}, \href
  {http://adsabs.harvard.edu/abs/1979SoPh...63..411M} {63, 411}

\bibitem[\protect\citeauthoryear{{Markwardt}}{{Markwardt}}{2009}]{markwardt09}
{Markwardt} C.~B.,  2009, in {Bohlender} D.~A.,  {Durand} D.,   {Dowler} P.,
  eds,  Astronomical Society of the Pacific Conference Series Vol. 411,
  Astronomical Data Analysis Software and Systems XVIII. p.~251 (\mn@eprint
  {arXiv} {0902.2850})

\bibitem[\protect\citeauthoryear{{Ogilvie} et~al.,}{{Ogilvie}
  et~al.}{1995}]{ogilvie95}
{Ogilvie} K.~W.,  et~al., 1995, \mn@doi [\ssr] {10.1007/BF00751326}, \href
  {http://adsabs.harvard.edu/abs/1995SSRv...71...55O} {71, 55}

\bibitem[\protect\citeauthoryear{{Ogilvie}, {Fitzenreiter}  \&
  {Desch}}{{Ogilvie} et~al.}{2000}]{ogilvie00}
{Ogilvie} K.~W.,  {Fitzenreiter} R.,   {Desch} M.,  2000, \mn@doi [\jgr]
  {10.1029/2000JA000131}, \href
  {http://adsabs.harvard.edu/abs/2000JGR...10527277O} {105, 27277}

\bibitem[\protect\citeauthoryear{{Pagel}, {Gary}, {de Koning}, {Skoug}  \&
  {Steinberg}}{{Pagel} et~al.}{2007}]{pagel07}
{Pagel} C.,  {Gary} S.~P.,  {de Koning} C.~A.,  {Skoug} R.~M.,   {Steinberg}
  J.~T.,  2007, \mn@doi [Journal of Geophysical Research (Space Physics)]
  {10.1029/2006JA011967}, \href
  {http://adsabs.harvard.edu/abs/2007JGRA..112.4103P} {112, A04103}

\bibitem[\protect\citeauthoryear{{Pavan}, {Vi{\~n}as}, {Yoon}, {Ziebell}  \&
  {Gaelzer}}{{Pavan} et~al.}{2013}]{pavan13}
{Pavan} J.,  {Vi{\~n}as} A.~F.,  {Yoon} P.~H.,  {Ziebell} L.~F.,   {Gaelzer}
  R.,  2013, \mn@doi [\apjl] {10.1088/2041-8205/769/2/L30}, \href
  {http://adsabs.harvard.edu/abs/2013ApJ...769L..30P} {769, L30}

\bibitem[\protect\citeauthoryear{{Pilipp}, {Muehlhaeuser}, {Miggenrieder},
  {Montgomery}  \& {Rosenbauer}}{{Pilipp} et~al.}{1987}]{pilipp87}
{Pilipp} W.~G.,  {Muehlhaeuser} K.-H.,  {Miggenrieder} H.,  {Montgomery} M.~D.,
    {Rosenbauer} H.,  1987, \mn@doi [\jgr] {10.1029/JA092iA02p01075}, \href
  {http://adsabs.harvard.edu/abs/1987JGR....92.1075P} {92, 1075}

\bibitem[\protect\citeauthoryear{{Pilipp}, {Muehlhaeuser}, {Miggenrieder},
  {Rosenbauer}  \& {Schwenn}}{{Pilipp} et~al.}{1990}]{pilipp90}
{Pilipp} W.~G.,  {Muehlhaeuser} K.-H.,  {Miggenrieder} H.,  {Rosenbauer} H.,
  {Schwenn} R.,  1990, \mn@doi [\jgr] {10.1029/JA095iA05p06305}, \href
  {http://adsabs.harvard.edu/abs/1990JGR....95.6305P} {95, 6305}

\bibitem[\protect\citeauthoryear{{Rathgeber} et~al.,}{{Rathgeber}
  et~al.}{2010}]{rathgeber10}
{Rathgeber} S.~K.,  et~al., 2010, \mn@doi [Plasma Physics and Controlled
  Fusion] {10.1088/0741-3335/52/9/095008}, \href
  {http://adsabs.harvard.edu/abs/2010PPCF...52i5008R} {52, 095008}

\bibitem[\protect\citeauthoryear{{Richardson} \& {Cane}}{{Richardson} \&
  {Cane}}{2010}]{richardsoncane10}
{Richardson} I.~G.,  {Cane} H.~V.,  2010, \mn@doi [\solphys]
  {10.1007/s11207-010-9568-6}, \href
  {http://adsabs.harvard.edu/abs/2010SoPh..264..189R} {264, 189}

\bibitem[\protect\citeauthoryear{{Rosenbauer} et~al.,}{{Rosenbauer}
  et~al.}{1977}]{rosenbauer77}
{Rosenbauer} H.,  et~al., 1977, Journal of Geophysics Zeitschrift Geophysik,
  \href {http://adsabs.harvard.edu/abs/1977JGZG...42..561R} {42, 561}

\bibitem[\protect\citeauthoryear{{Saito} \& {Gary}}{{Saito} \&
  {Gary}}{2007}]{saitogary07}
{Saito} S.,  {Gary} S.~P.,  2007, \mn@doi [\grl] {10.1029/2006GL028173}, \href
  {http://adsabs.harvard.edu/abs/2007GeoRL..34.1102S} {34, L01102}

\bibitem[\protect\citeauthoryear{{Salem}, {Hubert}, {Lacombe}, {Bale},
  {Mangeney}, {Larson}  \& {Lin}}{{Salem} et~al.}{2003}]{salem03}
{Salem} C.,  {Hubert} D.,  {Lacombe} C.,  {Bale} S.~D.,  {Mangeney} A.,
  {Larson} D.~E.,   {Lin} R.~P.,  2003, \mn@doi [\apj] {10.1086/346185}, \href
  {http://adsabs.harvard.edu/abs/2003ApJ...585.1147S} {585, 1147}

\bibitem[\protect\citeauthoryear{{Scudder} \& {Olbert}}{{Scudder} \&
  {Olbert}}{1979}]{scudderolbert79}
{Scudder} J.~D.,  {Olbert} S.,  1979, \mn@doi [\jgr] {10.1029/JA084iA11p06603},
  \href {http://adsabs.harvard.edu/abs/1979JGR....84.6603S} {84, 6603}

\bibitem[\protect\citeauthoryear{{Seough}, {Nariyuki}, {Yoon}  \&
  {Saito}}{{Seough} et~al.}{2015}]{seough15}
{Seough} J.,  {Nariyuki} Y.,  {Yoon} P.~H.,   {Saito} S.,  2015, \mn@doi
  [\apjl] {10.1088/2041-8205/811/1/L7}, \href
  {http://adsabs.harvard.edu/abs/2015ApJ...811L...7S} {811, L7}

\bibitem[\protect\citeauthoryear{{Smith}, {Marsch}  \& {Helander}}{{Smith}
  et~al.}{2012}]{smith12}
{Smith} H.~M.,  {Marsch} E.,   {Helander} P.,  2012, \mn@doi [\apj]
  {10.1088/0004-637X/753/1/31}, \href
  {http://adsabs.harvard.edu/abs/2012ApJ...753...31S} {753, 31}

\bibitem[\protect\citeauthoryear{{Spitzer} \& {H{\"a}rm}}{{Spitzer} \&
  {H{\"a}rm}}{1953}]{spitzerharm53}
{Spitzer} L.,  {H{\"a}rm} R.,  1953, \mn@doi [Physical Review]
  {10.1103/PhysRev.89.977}, \href
  {http://adsabs.harvard.edu/abs/1953PhRv...89..977S} {89, 977}

\bibitem[\protect\citeauthoryear{{Stansby}, {Horbury}, {Chen}  \&
  {Matteini}}{{Stansby} et~al.}{2016}]{stansby16}
{Stansby} D.,  {Horbury} T.~S.,  {Chen} C.~H.~K.,   {Matteini} L.,  2016,
  \mn@doi [\apjl] {10.3847/2041-8205/829/1/L16}, \href
  {http://adsabs.harvard.edu/abs/2016ApJ...829L..16S} {829, L16}

\bibitem[\protect\citeauthoryear{Vi{\~n}as, Gurgiolo, Nieves-Chinchilla, Gary
  \& Goldstein}{Vi{\~n}as et~al.}{2010}]{vinas10}
Vi{\~n}as A.,  Gurgiolo C.,  Nieves-Chinchilla T.,  Gary S.,   Goldstein M.,
  2010, Proc. Twelfth International Solar Wind Conference

\bibitem[\protect\citeauthoryear{{Vocks} \& {Mann}}{{Vocks} \&
  {Mann}}{2003}]{vocks03}
{Vocks} C.,  {Mann} G.,  2003, \mn@doi [\apj] {10.1086/376682}, \href
  {http://adsabs.harvard.edu/abs/2003ApJ...593.1134V} {593, 1134}

\bibitem[\protect\citeauthoryear{{Vocks}, {Salem}, {Lin}  \& {Mann}}{{Vocks}
  et~al.}{2005}]{vocks05}
{Vocks} C.,  {Salem} C.,  {Lin} R.~P.,   {Mann} G.,  2005, \mn@doi [\apj]
  {10.1086/430119}, \href {http://adsabs.harvard.edu/abs/2005ApJ...627..540V}
  {627, 540}

\bibitem[\protect\citeauthoryear{{Wilson} et~al.,}{{Wilson}
  et~al.}{2013}]{wilson13}
{Wilson} L.~B.,  et~al., 2013, \mn@doi [Journal of Geophysical Research (Space
  Physics)] {10.1029/2012JA018167}, \href
  {http://adsabs.harvard.edu/abs/2013JGRA..118....5W} {118, 5}

\bibitem[\protect\citeauthoryear{{de Koning}, {Gosling}, {Skoug}  \&
  {Steinberg}}{{de Koning} et~al.}{2007}]{dekoning07}
{de Koning} C.~A.,  {Gosling} J.~T.,  {Skoug} R.~M.,   {Steinberg} J.~T.,
  2007, \mn@doi [Journal of Geophysical Research (Space Physics)]
  {10.1029/2006JA011971}, \href
  {http://adsabs.harvard.edu/abs/2007JGRA..112.4101D} {112, A04101}

\bibitem[\protect\citeauthoryear{{{\v S}tver{\'a}k}, {Maksimovic},
  {Tr{\'a}vn{\'{\i}}{\v c}ek}, {Marsch}, {Fazakerley}  \& {Scime}}{{{\v
  S}tver{\'a}k} et~al.}{2009}]{stverak09}
{{\v S}tver{\'a}k} {\v S}.,  {Maksimovic} M.,  {Tr{\'a}vn{\'{\i}}{\v c}ek}
  P.~M.,  {Marsch} E.,  {Fazakerley} A.~N.,   {Scime} E.~E.,  2009, \mn@doi
  [Journal of Geophysical Research (Space Physics)] {10.1029/2008JA013883},
  \href {http://adsabs.harvard.edu/abs/2009JGRA..11405104S} {114, 5104}

\bibitem[\protect\citeauthoryear{{{\v S}tver{\'a}k}, {Tr{\'a}vn{\'{\i}}{\v
  c}ek}  \& {Hellinger}}{{{\v S}tver{\'a}k} et~al.}{2015}]{stverak15}
{{\v S}tver{\'a}k} {\v S}.,  {Tr{\'a}vn{\'{\i}}{\v c}ek} P.~M.,   {Hellinger}
  P.,  2015, \mn@doi [Journal of Geophysical Research (Space Physics)]
  {10.1002/2015JA021368}, \href
  {http://adsabs.harvard.edu/abs/2015JGRA..120.8177A} {120, 8177}

\makeatother
\end{thebibliography}

%\begin{thebibliography}{99}
%\end{thebibliography}

% Don't change these lines
\bsp	% typesetting comment
\label{lastpage}
\end{document}